\documentstyle[12pt,aasms4]{article}

\slugcomment{Accepted  to be published The Astrophysical Journal Supplement Series}

\lefthead{Gonz\'alez Delgado, Leitherer \& Heckman}
\righthead{H and HeI profiles: Evolutionary synthesis models}

\begin{document}

\title{Synthetic spectra of H Balmer and HeI absorption lines. II: Evolutionary synthesis models for starburst and post-starburst galaxies.}

\author{Rosa M. Gonz\'alez Delgado}
\affil{Instituto de Astrof\'\i sica de Andaluc\'\i a (CSIC), Apdo. 3004, 18080 Granada, Spain}
\affil{Electronic mail: rosa@iaa.es}

\author{Claus Leitherer}
\affil{Space Telescope Science Institute, 3700 San Martin Drive, Baltimore, MD
21218}
\affil{Electronic mail: leitherer@stsci.edu}

\and 

\author{Timothy M. Heckman}
\affil{The Johns Hopkins University, Department of Physics and Astronomy,
 Baltimore, MD 21218-2686}
\affil{Electronic mail: heckman@pha.jhu.edu}


\newpage

\begin{abstract}

We present evolutionary stellar population synthesis models to predict the spectrum 
of a single-metallicity stellar population, with a spectral sampling of 
0.3 \AA\ in five spectral regions between 3700 and 5000 \AA. The models, which are 
optimized for galaxies with active star formation, synthesize the profiles of the 
hydrogen Balmer series (H$\beta$, H$\gamma$, H$\delta$, H8, H9, H10, H11, H12 
and H13) and the neutral helium absorption lines (HeI $\lambda$4922, HeI $\lambda$4471,
HeI $\lambda$4388, HeI $\lambda$4144, HeI $\lambda$4121,  HeI $\lambda$4026,  
HeI $\lambda$4009 and HeI $\lambda$3819) for a burst with an age ranging from 
10$^6$ to 10$^9$ yr, and different assumptions about the stellar initial mass function 
(IMF). Continuous star formation models lasting for 1 Gyr are also presented. 
The input stellar library includes NLTE absorption profiles for stars hotter than 25000 K 
and LTE profiles for lower temperatures. The temperature and gravity coverage is 4000 K $\leq$T$_{eff}\leq$50000 K and 0.0$\leq$log g$\leq$5.0, respectively. 
The metallicity is solar.

It is found that the Balmer and HeI line profiles are sensitive to the age, except 
during the first 4 Myr of the evolution, when the equivalent widths of these lines 
are constant. At these early stages of the evolution, the profiles of the lines are 
also sensitive to the IMF. However, strong H Balmer and HeI lines are predicted even 
when the low mass cut-off of the IMF is as high as 10 M$\odot$. The equivalent widths 
of the Balmer lines range from 2 to 16 \AA\ and the HeI lines from 0.2 to 1.2 \AA.  
During the nebular phase (cluster younger than  about 10 Myr), H$\beta$ ranges from 2 to 
5 \AA\ and HeI $\lambda$4471 between 0.5 and 1.2 \AA. The strength of the lines is 
maximum when the cluster is a few hundred (for the Balmer lines) and a few ten 
(for the HeI lines) Myr old. In the continuous star formation scenario, the strength 
of the Balmer and HeI lines increases monotonically with time until 500 Myr and 100 Myr, respectively. However, the lines are weaker than 
in the burst  models due to the dilution of the Balmer and HeI lines by the 
contribution from very massive stars.

The high spectral resolution of the profiles is useful to reproduce the 
absorption wings observed in regions of recent star formation and to estimate the 
effect of the underlying absorption on the nebular emission-lines. The strength of the 
nebular emission Balmer and HeI lines compared with the stellar absorption 
components indicates that H$\delta$ and the higher-order terms of the Balmer series and 
HeI are dominated by the stellar absorption component if an instantaneous burst is older than $\simeq$ 5 Myr. Some of the HeI lines (e.g. HeI $\lambda$3819, HeI $\lambda$4922) are more favorable than others (e.g. HeI $\lambda$4471) for the detection of stellar features in the presence of nebular emission.
We estimate that the correction to the HeI $\lambda$4471 nebular emission line due 
to the stellar absorption is between 5 and 25$\%$, if the nebular emission has equivalent 
width between 10 and 2.5 \AA\ (corresponding to a burst age between 1-3 Myr).

The models can be used to date starburst and post-starburst galaxies until 1 Gyr. 
They have been tested on data for clusters in the LMC, the super-star cluster B in 
the starburst galaxy NGC 1569, the nucleus of the dwarf elliptical NGC 205 and a luminous "E+A" galaxy. The full data set is 
available for retrieval at http://www.iaa.es/ae/e2.html and at http://www.stsci.edu/science/starburst/, 
or on request from the authors at rosa@iaa.es

\end{abstract}



\newpage
\section{Introduction}

Starburst galaxies are characterized by a nebular emission-line spectrum at 
optical and an absorption-line spectrum at ultraviolet wavelengths. This dichotomy arises because starbursts are powered by massive 
stars (M$\geq$ 10 M$\odot$). These stars emit photons with energies of a few 
tens of eV which are absorbed and re-emitted in their stellar winds, producing 
ultraviolet resonance transitions, and thus, an ultraviolet absorption line spectrum.
However, most of the ionizing photons emitted by the stars can travel tens of pc 
or more before they are absorbed by the surrounding interstellar gas. Then, this ionized gas 
emits a nebular emission-line spectrum, mainly at optical and near-infrared wavelengths. However, 
around the Balmer jump the spectra can show absorption features formed in the photospheres
of massive stars.

The radiative properties of starburst galaxies are determined by their massive stellar content.
O, B and A stars can dominate the optical continuum emission of starburst galaxies.
The spectra of the early-type stars are characterized by strong hydrogen Balmer and 
neutral helium absorption lines and with only very weak metallic lines (Walborn \&
Fitzpatrick 1990). However, the detection of these stellar features at optical wavelengths
in the spectra of starburst galaxies is difficult because H and HeI absorption
features are coincident with the nebular emission-lines that mask the absorption features.
Even so, the higher-order terms of the Balmer series and some of the HeI lines are
detected in absorption in many starburst galaxies (Storchi-Bergmann, Kinney, \& Challis 1995; 
Gonz\'alez-Delgado et al. 1995), in Seyfert 2 galaxies with their ultraviolet and optical 
continuum dominated by a nuclear starburst (Shields \& Filippenko 1990; Gonz\'alez Delgado et al. 1998;
Cid Fernandes, Storchi-Bergmann, \& Schmitt 1998), or even in the spectra of giant 
HII regions (e.g. NGC 604, Terlevich et al. 1996). These features can be seen in absorption 
because the strength of the Balmer series in emission decreases rapidly with decreasing 
wavelength, whereas the equivalent width of the stellar absorption lines is constant 
or increases with wavelength. Thus, the net effect is that H$\alpha$, H$\beta$ and H$\gamma$
are mainly seen in emission, but the higher-order terms of the series are seen 
in absorption. Very often, H$\alpha$, H$\beta$, H$\gamma$ and H$\delta$ show absorption wings
superposed on the nebular emission. A similar effect occurs for the HeI lines. The strongest 
nebular emission occurs in the HeI $\lambda$5876 and HeI $\lambda$6678 whereas the stellar 
absorption features are very weak. Meanwhile, the equivalent width of the nebular emission 
of HeI $\lambda$4471 or HeI $\lambda$4026 ($\leq$ 6 \AA,  Izotov, Thuan, \& Lipovetsky 1997) is similar or even smaller than that of the stellar absorption lines. Thus, they can be easily detected in absorption.

Models of the H Balmer series and HeI lines of a stellar population in a starburst galaxy can be used to predict
the evolutionary state of a starburst, the effect of the initial mass function (IMF) and corrections to the nebular H and HeI emission-lines. Accurate measurements of the nebular emission 
lines are needed to estimate the amount of internal interstellar reddening, star formation rate, 
excitation parameter and chemical abundances (in particular in the determination of the primordial helium abundance) in starburst galaxies.
All these estimations are affected by the underlying H and HeI stellar absorption. The internal 
interstellar reddening is derived using the Balmer decrement. The emission-line ratios
(H$\alpha$/H$\beta$, H$\gamma$/H$\beta$, H$\delta$/H$\beta$, ...) are clearly affected by the stellar absorption. They are not affected equally because nebular emission decreases 
with wavelength. Reddening and underlying absorption affect line ratios such as 
[OII] $\lambda$3727+ [OIII] $\lambda$5007/H$\beta$ and [OIII] $\lambda$5007/H$\beta$
which are used to estimate chemical abundances and the excitation parameter in 
starburst galaxies. The strength of the collisionally-excited emission-lines with respect to
H$\beta$ can be used to estimate the evolutionary state of starburst galaxies 
(Garc\'\i a-Vargas, Bressan, \& D\'\i az 1995; Garc\'\i a-Vargas et al. 1997; 
Gonz\'alez Delgado et al. 1999); higher line ratios are derived if H$\beta$ is not 
corrected for the stellar absorption. The underlying stellar absorption in the HeI lines leads 
to an understimate of the He abundance in starburst galaxies, and potentially of the determination of the primordial helium abundance. However, high resolution 
spectra are needed to isolate the absorption lines from the nebular emission.

On the other hand, the strength of the higher-order lines of the Balmer series and 
some of the HeI lines (e.g. HeI $\lambda$4026, $\lambda$3819), which are likely to be detected in absorption in
starburst galaxies can be used to obtain information about the age and IMF of their stellar
population. It has been suggested that starburst galaxies have a deficit of intermediate and low-mass stars
(Rieke et al. 1980; Viallefond \& Thuan 1983; Augarde \& Lequeux 1985; Scalo 1990). However, the evidence 
in favor of this top-heavy IMF in starburst galaxies is ambiguous (Satyapal et al 1995,1997). It should be expected that the strength of the Balmer absorption lines depends on the 
low mass cut-off because these lines are stronger in A stars. Thus, if the mass of the starburst
is distributed following a top-heavy IMF, we expect that the H Balmer
lines are weaker than for the population with a non-truncated Salpeter IMF.

H Balmer and HeI lines of a stellar population can be predicted using the 
evolutionary synthesis technique, a powerful tool to derive the properties 
of a stellar population taking as a free parameter the star formation history of the
starburst (age, IMF, star formation rate). This technique has been used previously to derive the strength of 
the Balmer and He I absorption lines of star forming regions. D\'\i az (1988) used the synthetic 
profiles of H$\alpha$, H$\beta$, H$\gamma$ and H$\delta$ computed by Kurucz (1979).
She followed the evolution of the stellar cluster up to 40 Myr and found that the 
profiles are rather insentive to the age and IMF. Olofsson (1995) computed the equivalent
widths of the hydrogen (H$\alpha$, H$\beta$, H$\gamma$, H$\delta$ and L$\alpha$) and 
HeI ($\lambda$4026, 4387, 4471 and 4921) lines of a single-burst star forming regions at three 
different metallicities ($\it Z$)=0.001, 0.008 and 0.020), and followed the evolution up to 15 Myr.  He used the synthetic models of Kurucz (1993) and Auer \& Mihalas (1972) at a resolution of 5 \AA.
He found that the equivalent width of the Balmer lines increases during the course
of the evolution, while HeI decreases monotonically. The strength of these lines are
very sensitive to the IMF and insensitive to the metallicity. He suggested that 
Balmer lines are prominent even if the low mass cut-off is as high as 10 M$\odot$. Cananzi,
Augarde \& Lequeux (1994) predict the Balmer lines (H$\beta$, H$\gamma$ and H$\delta$)
using stellar spectra of O to G type observed at a spectral resolution of 2 \AA\ (Cananzi, 
Augarde, \& Lequeux 1993). They conclude that the equivalent widths of the lines are
more sensitive to the low than to the high-mass cut-off of the IMF. 

The results seem contradictory: it is not yet clear if
the IMF parameters have an effect on the strength of the H Balmer and HeI lines.
On the other hand, none of the previous studies made predictions for the strength of the higher-order Balmer series lines (H8 to H13) which are least affected by the 
emission and the most useful to obtain information on the evolutionary state of the 
starburst. Due to the coincidence of the nebular emission and absorption lines in the 
observed spectra of starbursts, models that compute the profiles of these lines at 
high spectral resolution are required to predict the underlying absorption through 
the analysis of the absorption wings. To archieve this
 goal, we have built a stellar library of synthetic spectra with a sampling of 0.3 \AA, 
effective temperature from 50000 to 4000 K and gravity between 
0.0$\leq$log g$\leq$5.0 (Gonz\'alez Delgado \& Leitherer 1999, paper I).  
The profiles are computed using a set of programs developed by Hubeny and 
collaborators (Hubeny 1988; Hubeny \& Lanz, 1995a,b; Hubeny, Lanz, \& Jeffery, 1995) 
in three stages. For T$_{eff}\geq$ 25000 K, the code TLUSTY is used to compute NLTE 
stellar atmosphere models. These models together with Kurucz (1993) LTE stellar 
atmosphere models (for T$_{eff}\leq$ 25000 K) are used as input to SYNSPEC, the 
program that solves the radiative transfer equation. Finally, the synthetic spectrum is obtained after performing the rotational and instrumental convolution. This library has been included in our evolutionary synthesis code (Starburst99, Leitherer et al 1999) 
to predict the strength of the Balmer and He I lines for a stellar population.  

This paper is organized as follows: Section 2 describes the models, the assumptions and 
computational techniques. Section 3 deals with the results and discusses the effects of the IMF 
parameters and metallicity on the strength of the lines.  We compare the equivalent widths of the Balmer and HeI absorption lines with the values predicted for the nebular recombination emission-lines in Section 4. In Section 5,  
the models are compared with observations of stellar clusters of the Large Magellanic Clouds, the super-star clusters in the starburst galaxy NGC 1569, the nucleus of the dwarf elliptical galaxy NGC 205 and a luminous "E+A" galaxy. The summary and the conclusions are in Section 6.

\section{Description of the models}

The stellar library of synthetic spectra which covers the main hydrogen Balmer 
and neutral He lines (paper I), has been implemented in our evolutionary synthesis code 
Starburst99 (Leitherer et al. 1999). The synthesis is done in two steps. First, 
the code computes the population of stars in the cluster as a function of age
and IMF; then the profiles of the lines are synthesized by adding the different
contributions from stars.

Two limiting models of the star formation history are considered: an instantaneous burst
and star formation proceeding continuously at a constant rate.
In the instantaneous burst, all the stars are formed at the same time on the zero-age-main-sequence (ZAMS),
and then evolve in the Hertzsprung-Russell diagram (HRD) until 1 Gyr. These models 
are normalized to a total mass of 10$^6$ M$\odot$. The star formation rate of the 
continuous model is 1 M$\odot$ yr$^{-1}$; they extend to 1 Gyr. 

The models are generated for different assumptions about the IMF, which is 
parameterized as a power law, $\Phi$(m)= C m$^{-\alpha}$; where the constant $C$
is determined by the total gas mass converted into stars. The reference model is a 
power law with exponent $\alpha$=2.35, and with low-mass and high-mass cut-off
masses of M$_{low}$= 1 M$\odot$ and M$_{up}$= 80 M$\odot$. For comparison, 
we have computed models with $\alpha$= 3.3 and $\alpha$= 1.0 between 
$M_{low}$= 1 M$\odot$ and $M_{up}$= 80 M$\odot$; and a  truncated Salpeter 
($\alpha$ = 2.35) IMF between $M_{low}$= 1 M$\odot$ and $M_{up}$= 30 M$\odot$, $M_{low}$= 5 M$\odot$ and $M_{up}$= 80 M$\odot$, and $M_{low}$= 10 M$\odot$ and $M_{up}$= 80 M$\odot$.

The models are computed with the isochrone synthesis method. Thus, continuous
isochrones are calculated by interpolating between the tracks in the HRD on a
variable mass grid. The stars evolve from the ZAMS following the new set of 
evolutionary tracks of the Geneva group (Schaller et al. 1992, Schaerer 
et al. 1993a,b, Charbonnel et al. 1993, Meynet  et al. 1994), with standard-mass
loss rate. 

The time resolution of the models is 0.01 Myr; however, 
the profile of the lines are given only at time steps during which significant
 changes occur. For every step, the luminosity, effective temperature, radius
and surface gravity are calculated for each star. Then a spectrum of our stellar 
library is assigned to each star. The corresponding synthetic spectrum is flux 
calibrated using the stellar atmosphere grid compiled by Lejeune et al (1997).
This compilation which includes the 20 \AA\ resolution Kurucz (1993) stellar 
atmospheres is supplemented by Schmutz, Leitherer, \& Gruenwald (1992) models for
stars with expanding envelopes. The flux of the continuum is assigned to each 
normalized spectrum by fitting a first order polynomial 
to the plane (F$_\lambda$, log $\lambda$), where F$_\lambda$ is the distribution 
of the continuum flux of the corresponding low resolution stellar atmosphere model. 
The fit is done taking 
points of the continuum adjacent to the absorption lines H$\beta$, H$\gamma$, H$\delta$ 
and HeI $\lambda$4471 for the spectral ranges 4820-4950 \AA, 4270-4430 \AA, 3990-4150 \AA\ and 4420-4580 \AA, 
respectively. To 
calibrate the stellar spectra between 3720 and 3920 \AA, we use the result of 
the fit in the wavelength range of H$\delta$ (if T$_{eff}\geq$ 12000 K) or 
H$\beta$ (if T$_{eff}\leq$ 12000 K) extrapolated to those wavelengths. The
calibration was chosen to achieve consistency between the distribution of the 
continuum flux of our synthesized high-resolution spectra and the spectral
energy distribution which is also predicted by our code at a resolution of 20 \AA\ (see 
Figure 1). The nebular continuum has been added to the synthesized spectra. 
We have assumed that all the stellar far-UV photons below 912 \AA\ are 
absorbed by gas and converted into continuous and line emission at longer wavelengths.

We have computed the synthesized profiles for two metallicities, $\it Z$=0.02 (Z$\odot$) and
Z=0.001 (1/20 Z$\odot$). While the input stellar evolutionary tracks are different for these two metallicities, we use the same stellar library (generated assuming solar 
metallicity) because the profiles of the Balmer lines of individual stars do not change 
significantly with metallicity for T$_{eff}\geq$ 7000 K (see paper I). 
 
\section{Model results}

Figure 2 shows the synthetic spectra for a star cluster formed instantaneously
3, 50 and 500 Myr ago. The IMF is Salpeter with $M_{low}$= 1 M$\odot$
and $M_{up}$= 80 M$\odot$; the metallicity is solar. The general tendency is that 
the strength of the Balmer lines increases with the evolution until 500 Myr, and 
the HeI lines until 30-40 Myr. 

The equivalent widths of the Balmer (Table 1 and 2) and He I (Table 3 and 4) lines 
are measured in the windows indicated in the Figure 2. Two different measurements are made, 
one integrating the flux from the continuum set to unity and the other from a 
pseudo-continuum. This pseudo-continuum is found fitting a first-order polynomial
(except for the spectral range 3720-3920 \AA, for which we use a third-order fit)
to the continuum windows defined in Table 3 of paper I.  The absolute values of the equivalent
widths depend on the continuum level and on the width on the windows used to measure the strength
of the lines. However, the relative behavior of the equivalent width with age of the star cluster is 
independent of the continuum and of the width of the windows. We plot the equivalent widths
measured from the pseudo-continuum because these measurements are more representative of
the equivalent widths we can measure in observed spectra. 

First, we discuss the profiles and equivalent widths of H Balmer and He I lines for the reference 
model (Salpeter IMF, M$_{low}$= 1 M$\odot$ and M$_{up}$= 80 M$\odot$) for an 
instantaneous burst and for continuous star formation. Then, the effect of the IMF 
parameters and finally the effect of the metallicity are discussed. The full data set is available 
for retrieval at http://www.iaa.es/ae/e2.html and at http://www.stsci.edu/science/starburst/.

\subsection{The hydrogen absorption lines}

Figure 3 shows the equivalent widths of the Balmer lines as a function of the age for an 
instantaneous burst. We plotted the equivalent width measured from the pseudo-continuum 
and integrating the flux in the windows of 60 \AA\ for H$\beta$, H$\gamma$ and H$\delta$, 
and 30 \AA\ for H8, H9 and H10, respectively. All the Balmer lines have similar behavior; 
the equivalent widths are in the 
range between 2 \AA\ to 16 \AA. During the first 4 Myr, the equivalent width does not change, 
with H$\beta$, H$\gamma$ and H$\delta$ $\sim$ 3 \AA, which is very similar to the values 
predicted by Olofsson (1995), and very similar to the value estimated ($\sim$ 2 \AA) 
for the Balmer absorption in giant HII regions (Shields \& Searle 1978, McCall, 
Ribsky, \& Shields 1985). 

The equivalent width increases with the age of the burst until 500 Myr; then, it 
decreases. The maximum is at 500 Myr since the turn-off stellar population in 
the cluster are stars of type A2V that have a life-time of several hundreds Myr. These 
stars have T$_{eff}$ of 9000 K, and at this T$_{eff}$ the strength of the Balmer 
lines peaks (see e.g. Figure 2 in paper I). The predicted value at 500 Myr (the equivalent width 
of H$\beta$= 11-12 \AA) is, e.g., in agreement with the values measured in the super star 
clusters of NGC 1275 (Brodie et al. 1998) and the values predicted by population synthesis 
models using stellar clusters (Bica \& Alloin 1986a).

For continuous star formation, the strength of the lines increases with time until 500 Myr; then, the equivalent widths are constant (Figure 4). 
The equivalent widths range
from 3 to 10 \AA. They are very similar to the burst models at the begining of the 
evolution (about 3 \AA\ for continuous star formation lasting for 4 Myr). However, 
the equivalent width increases less rapidly with time than in the burst model. 
The reason is the dilution of the Balmer lines by O stars which have weaker Balmer lines 
than A stars and which are continuously forming in the cluster. 

\subsection{The neutral He absorption lines}

Figure 5 shows the equivalent width of some of the HeI lines. HeI $\lambda$4471 and 
HeI $\lambda$4026 are the strongest lines; they have equivalent widths that range 
between 0.5 \AA\ and 1.1 \AA. HeI $\lambda$4922, HeI $\lambda$4388 and 
HeI $\lambda$3819 are about a factor two weaker than HeI $\lambda$4471 and HeI $\lambda$4026. 
This result is opposite to that found by Olofsson (1995). However, observations of 
individual Galactic B stars (Lennon, Dufton, \& Fitzsimmons 1993) show that the lines 
HeI $\lambda$4471 and HeI $\lambda$4026  have
both similar strength and are about a factor of two stronger than 
HeI $\lambda$4922 and HeI $\lambda$4388 (these two lines have also similar strength). 
We found the same result examining the spectral atlas of Walborn \& Fitzpatrick (1990). 

The equivalent width of the HeI lines is almost constant for bursts younger than 4 Myr, with HeI $\lambda$4471 and HeI $\lambda$4026 being about 0.5 \AA, and HeI $\lambda$4922, 
HeI $\lambda$4388 and HeI $\lambda$3819 about 0.3 \AA. After 4 Myr, the equivalent 
width increases with age (but not monotonically) until the burst age reaches 30-40 Myr due 
to the contribution of early-type B stars that show the strongest HeI 
absorption. This result is opposite to that found by Olofsson (1995). He followed the 
evolution of the starburst until 15 Myr, and he found that the maximum strength of the 
HeI lines is at 2 Myr, no HeI absorption is formed in his models burst older than 10 Myr. 
Lennon et al (1993) show that stars from O9 to B2 have
similar strengths of the HeI $\lambda$4471 and HeI $\lambda$4026 lines; these are weaker 
in stars hotter than O9 and cooler than B2. By definition HeI lines are not formed in the 
photosphere of stars cooler than B9.
These observations are in agreement with the results found in our stellar grid 
(see Figure 4 of paper I) and  explain why the equivalent width of these lines in the 
integrated light of a stellar cluster have their maximum values between 10 and 40 Myr, 
when the cluster is dominated by early B stars. HeI absorption is not formed for bursts 
older than 100 Myr because the life-time of B9 stars is about this age.

The equivalent width has some local maximum between 5 and 20 Myr. This is due to the 
presence of post-main sequence stars with T$_{eff}\leq$8000 K that contribute significantly 
to the integrated light of the cluster at these ages (Figure 6). These are the blue loops of the red super-giant in the tracks. These stars do not produce 
HeI lines but show many metal lines close to the HeI lines (Figure 7). These 
metal lines are in the spectral windows used to measure the equivalent width. They depress 
the continuum and contribute to the integrated profile making it broader and asymmetric 
(Figure 8). A similar effect is also seen in the H Balmer lines. However, the contribution 
of these metallic lines to the equivalent width of the window where the Balmer lines are 
measured is less than to the windows of the HeI lines because the Balmer lines are relatively 
stronger.

For continuous star formation, the equivalent width ranges between 0.5 to 0.9 \AA\ 
(for HeI $\lambda$4471 and HeI $\lambda$4026). The strength of these lines increases 
with time until $\simeq$ 100 Myr; then, the equivalent widths decrease with time (Figure 9). 40 Myr after the onset of the star formation, 
the equivalent widths of the lines (HeI $\lambda$4471 and HeI $\lambda$4026) is 0.8 \AA. 
They are weaker than in the instantaneous burst model of the same age. However, due to 
the presence of B stars in the cluster, the HeI lines are stronger in the 
continuous star formation scenario than in the instantaneous burst after 100 Myr.

\subsection{The hydrogen Balmer and HeI absorption lines as a function of the IMF 
parameters}

We study the influence of the slope and the low and high mass cut-off of the IMF on the profile 
and equivalent width of the hydrogen Balmer and HeI lines in a single stellar population. 
The models computed have IMF  slopes $\alpha$=
1.0 and $\alpha$= 3.3, low mass cut-offs $M_{low}$= 5 M$\odot$ and $M_{low}$= 10 M$\odot$ 
and high mass cut-offs $M_{up}$= 30 M$\odot$ with respect to the reference model 
($\alpha$= 2.35, $M_{low}$= 1 M$\odot$ and $M_{up}$= 80 M$\odot$).

Figure 10 shows the change of the profiles of H$\delta$ and HeI ($\lambda$4026, 4009 
and 4143) with the slope of the IMF for a 2 Myr old burst. Clusters formed with an IMF
flatter than Salpeter produce hydrogen Balmer lines weaker than in the reference 
model because more massive stars are formed in the cluster. At 2 Myr, these stars dilute 
the strength of the Balmer and HeI lines since these lines are weaker in O than in B stars.
However, after 4 Myr the relative number of massive stars still on the main
sequence with respect to B and A stars is very low. At that epoch, the profiles and equivalent widths 
of these lines are almost the same as those of the reference model (Figure 11).
HeI lines are also weaker than in the reference model if the cluster is younger than
4 Myr. These lines are slightly stronger if the cluster is between 20 and 100 Myr old
(Figure 12). During the post-starburst phase, the integrated light is dominated by B stars, 
and for a flatter IMF, the relative number of B stars
with respect to A stars is slightly higher than in the reference model. This 
effect, however, is very small.
The opposite result is found if the IMF is steeper than the Salpeter IMF. In this case the Balmer 
and He I lines are stronger than in the reference model if the burst is younger than
4 Myr. Between 20 and 100 Myr, HeI lines are weaker than in the Salpeter model, because 
fewer B stars contribute to the integrated light of the cluster. However, at these ages the change 
of the strength of these lines (Figure 12) is less important than at ages younger than 4 Myr,
when the equivalent width can change by almost a factor of 2 (Figure 11 and 12).

Figure 13 shows the change of H$\delta$ and HeI ($\lambda$4026, 4009 
and 4143) with $M_{low}$ and $M_{up}$ for a 2 Myr old burst. The H$\delta$ 
and HeI lines are stronger if $M_{up}$= 30 M$\odot$ than in the reference model; 
but they are very similar to those predicted by the reference model if 
$M_{low}$= 5 M$\odot$. This result holds as long as the burst is younger than 5 Myr 
because the life-time of stars with mass higher than 30 M$\odot$ is less than 5 Myr. Thus, if 
the cluster is younger than 5 Myr and the mass is distributed following a standard 
Salpeter IMF, the integrated light should be dominated by stars more massive than 30 M$\odot$.
These stars have Balmer and HeI lines that are weaker than stars less massive than
30 M$\odot$. In this case, the Balmer and HeI lines are diluted with respect to the truncated 
($M_{up}$=30 M$\odot$) IMF. However, after 5 Myr, all the stars more massive than
30 M$\odot$ have evolved from the main sequence and the integrated light shows 
Balmer and HeI lines of the same strength. 

If the cluster is formed following a truncated Salpeter IMF with $M_{low}$=5 M$\odot$,
the integrated light shows the Balmer and HeI lines at a similar strength as in a cluster
formed following the standard Salpeter IMF. The life-time of stars with mass of 5 M$\odot$
is about 100 Myr. Clusters formed without stars less massive than 5 M$\odot$ will
fade after 100 Myr when all the stars have evolved from the main-sequence. A stronger effect
of a truncated IMF is found if $M_{low}\geq$ 5 M$\odot$. In these cases, Balmer and HeI lines
are always weaker than in the reference model because fewer  B and A stars are formed.

The main effect of changing the parameters of the IMF is observed in the first 4 Myr of the evolution 
of the starburst. If the IMF is steeper than the reference one or the upper mass cut-off is much
lower than 80 M$\odot$, fewer O stars are formed and the Balmer and HeI lines
are stronger. If the IMF is flatter than Salpeter or if the low mass cut-off is
higher than 5 M$\odot$, more O stars are formed relative to B and A stars, and the lines are weaker.
However, the ability of the Balmer and HeI lines to constrain the IMF depends strongly on the 
resolution of the observations, and on the spatial distribution of the stellar cluster with respect 
to the nebular emission, because at ages younger than 5 Myr starbursts show very strong 
HeI and Balmer emission-lines that fill in the absorptions. 
A nearby starburst, that can be spatially resolved, and where the core stellar cluster shines through 
a hole of nebular emission produced by the stellar winds, will show a spectrum with absorption wings
around the Balmer emission and with the HeI lines in absorption; NGC 604 is a good example of this
(see Figure 9 in Terlevich et al 1996). In this cases, the HeI lines that are fainter in emission 
(e.g., HeI $\lambda$4026, $\lambda$4388 and $\lambda$4026) can be dominated by the stellar absorption. 
Thus, under these circumstances, the equivalent widths of these lines and the wing absorption profiles 
of the higher terms of the 
Balmer series can be used to constrain the IMF (Gonz\'alez Delgado \& P\'erez 1999, in preparation). 
When these circumstances do not occur, the total integrated spectrum of a starburst will be dominated 
by the nebular emission while it is younger than 5 Myr (see also section 4), and the stellar lines will not be 
detected; therefore, any information about the starburst will be obtained through the analysis 
of the emission-lines. 
   
\subsection{The effect of metallicity}

In paper I we demonstrated that the strengths of the Balmer and HeI lines in {\em individual} 
stars hotter than 7000 K do not depend on metallicity. However, the strength of 
these lines in the {\em integrated} light of a stellar cluster depends on metallicity 
due to the metallicity dependent stellar population. At low metallicity, a star of a given ZAMS mass is hotter and evolves more 
slowly in the HRD diagram, explaining why the Balmer lines are weaker at Z= 0.001 than 
at solar metallicity (Figure 14, Table 5). At any age younger than 500 Myr, main-sequence stars 
that dominate the luminosity of the cluster have higher 
effective temperature at  Z= 0.001 than at solar metallicity. After 500 Myr, 
stars hotter than 9000 K are still in the main sequence at Z= 0.001 metallicity. Therefore, 
the equivalent width of the Balmer lines can still increase with the age of 
the cluster. In contrast, at solar metallicity, main-sequence stars have effective temperature 
lower than 9000 K. As a result, the equivalent width decreases with age. 

The effect is similar for the HeI lines. For ages younger than 20 Myr, the lines are 
weaker at  Z= 0.001 than at solar metallicity, but between 50-300 Myr there are 
still main-sequence stars with effective temperature hotter than 12000 K that 
contribute significantly to the formation of the HeI lines. Therefore, the equivalent width 
of the HeI lines in this age range is higher at Z= 0.001 than at solar metallicity (Figure 15, Table 6).  

\section{H Balmer series and HeI stellar absorption {\it vs} nebular emission-lines}

In this section, we compare the equivalent width of the nebular H Balmer series and HeI lines and the 
stellar absorption lines. The goal is to quantify the effect of the stellar absorption on the strength 
of the nebular lines and determine the evolutionary stages of the starburst during which this effect is important. 
This will also show which of the Balmer and HeI lines are more useful to age-date starbursts. 

We calculate the equivalent width of the H Balmer and HeI nebular emission-lines as the ratio of 
the luminosity of the emission-line to the continuum, which is estimated from the continuum 
adjacent to the line and includes the luminosity of the stellar and nebular continuous emission. 
The luminosities of the H Balmer nebular lines is calculated from the relationship between the 
intensity of the line and the number of Lyman continuum photons and the theoretical ratios 
(assuming case B; Osterbrock 1989) H$\gamma$/H$\beta$ (0.469), H$\delta$/H$\beta$ (0.259) 
and H8/H$\beta$ (0.105). The luminosity of the HeI recombination lines have been calculated 
from the ratio HeI $\lambda$4471/H$\beta$ and the theoretical ratio  
HeI $\lambda$4026/HeI $\lambda$4471 (0.474), HeI $\lambda$4922/HeI $\lambda$4471 (0.274) 
and HeI $\lambda$3819/HeI $\lambda$4471 (0.264). 

We estimate the ratio HeI $\lambda$4471/H$\beta$ using the photoionization code CLOUDY 
(Ferland 1997). The models are computed assuming that the gas is ionization bounded and is spherically distributed 
around the ionizing cluster with a constant density of 100 cm$^{-3}$. The chemical composition 
of the gas is solar and the ionizing photon luminosity is fixed to the values predicted by the 
evolutionary synthesis models. We take as the radiation field the spectral energy distribution 
predicted by Starburst99 (Leitherer et al. 1999). Models are computed for an instantaneous burst 
with age between 0 and 20 Myr and continuous star formation lasting up to 100 Myr. Tables 7 
and 8 show the ratio HeI $\lambda$4471/H$\beta$ and the equivalent width 
of the H and HeI lines for burst and continuous star formation, respectively.

The ratio HeI $\lambda$4471/H$\beta$ decreases with the age of the instantaneous burst, 
and drops to zero for a burst older than 5 Myr. These clusters do not harbor stars that produce a significant number of photons to ionize He into He$^+$; therefore, the ratio should be zero. Csf models lasting for 
more than 5 Myr predict a HeI $\lambda$4471/H$\beta$ ratio which is constant with time, 
because after this age the number of stars able to ionize He is constant since births and deaths 
are in equilibrium. The stellar absorption is more important for the higher terms of the Balmer 
series and for older instantaneous bursts than for the csf scenario (Figure 16). H$\beta$ is dominated 
by the stellar absorption if the burst is older than 8 Myr; however, the effect of the stellar 
absorption is small if the star formation proceeds continuously. Thus, at 100 Myr, the strength 
of the H$\beta$ stellar absorption is less than 20$\%$ of the strength of the nebular emission. 
However, the effect of the stellar absorption is more dramatic for the higher terms of the Balmer series. H$\delta$ and H8 are dominated by the stellar absorption if the burst is older than 4 and 5 Myr, 
respectively. If the star formation proceeds continuously, after 30 Myr and 100 Myr, the strengths 
of the stellar absorption lines are equal to those of the nebular emission-lines H8 and H$\delta$, respectively.

The HeI nebular emission-lines are more affected by the stellar absorption lines than the hydrogen Balmer lines  
(Figure 17). Bursts older than 5 Myr do not show HeI in emission, and after this age they can be in 
absorption (Figure 17a). HeI $\lambda$3819 is more affected by the stellar absorption than 
the other lines. The equivalent width of the emission-line equals the absorption equivalent width 
when the burst is only 3 Myr old. As with the Balmer lines, HeI lines are less affected by the absorption 
if the star formation proceeds continuously (Figure 17b). HeI $\lambda$4471 is very little affected 
by the absorption; however, the equivalent width of the emission-line HeI $\lambda$3819 equals 
the stellar absorption line equivalent width at 20 Myr for continuous star formation, and after this time HeI $\lambda$3819 is dominated by the stellar absorption.  

Thus, as we anticipated, the high-order terms of the Balmer series and HeI $\lambda$3819 in 
absorption are good indicators of the evolutionary state of the starburst, because the nebular 
emission is weaker and the lines are dominated by the stellar component.

\section{Testing the model results}

In this section, we compare the profile and 
equivalent width of the Balmer lines of our models with observations of stellar 
clusters in the Large Magellanic Cloud (LMC), the super-star cluster B in the starburst galaxy NGC 1569, the nucleus 
of the dwarf elliptical galaxy NGC 205, and a luminous "E+A" galaxy to test the applicability of our models.

\subsection{Clusters in the LMC}

Bica \& Alloin (1986b) observed clusters in the LMC with ages between 10 and 500 Myr. 
The data were grouped in four different types, called Y1, Y2, Y3 and Y4, corresponding to ages of 
10 Myr, 25 Myr, 80 Myr and 200-500 Myr, respectively. These spectra are public (Leitherer et al 1996). 
The spectral range covers 1200 to 9800 \AA, with a resolution  in the range 7-17 \AA. At optical 
wavelengths, the template Y1 contains the cluster NGC 2004 (age= 8 Myr, [Z]=-0.25); 
Y2 contains NGC 1847 (age= 25 Myr, [Z]=-0.4), NGC 2157 (age= 30 Myr, [Z]=-0.6) 
and NGC 2214 (age= 40 Myr, [Z]=-1.2); Y3 contains NGC 1866 
(age= 86 Myr, [Z]=-1.2); and Y4 contains NGC 1831 (age= 300 Myr, [Z]=-1.0) 
and NGC 1868 (age= 500 Myr, [Z]=-1.1). 

Figure 18 compares the equivalent width of the Balmer lines of the four templates 
measured in the same windows as the models. The figures show the model results for a single 
burst assuming that the mass of the cluster is distributed with a Salpeter IMF between 
1 and 80 M$\odot$ at Z$\odot$ (full line) and Z= 0.001 (dashed line) metallicity. 
The observations and the models are in very good agreement. However, the equivalent 
widths of the higher terms of the Balmer series of the template Y4 (age=200-500 Myr) 
are weaker than in the models (Figure 18 c y d). The discrepancy could be related to the poor 
spectral resolution of the observations. Decreased spectral resolution artificially decreases the 
continuum level; thus, decreasing the equivalent width of the observed lines. This effect 
is more important for template Y4 because the Balmer lines are stronger and broader for ages 
between 200 and 500 Myr.

\subsection{Super-star cluster B in NGC 1569}

Super-star clusters (SSC) are highly compact (diameter a few pc), massive 
(10$^5$-10$^6$ M$\odot$), young (age younger than 1 Gyr) clusters. 
There are suggestions that SSCs represent the present-day analogs of young 
globular clusters because their masses and sizes are comparable to those of 
Milky Way globular clusters (O'Connell, Gallagher, \& Hunter 1994; Ho \& Filippenko 1996). 
They represent a basic mode of star formation in starburst galaxies (Meurer et al. 1995) and 
are building blocks of these objects. The clusters can be approximated by a coeval, single metallicity stellar population.

HST images of the dwarf starburst galaxy NGC 1569 suggest that the SSC B has an age of 15-300 Myr 
(O'Connell et al. 1994) and ground-based optical spectra suggest an age of 10 Myr 
(Gonz\'alez Delgado et al. 1997). The latter result is based on the analysis of the 
optical spectral energy distribution and on the strength of the CaII triplet 
(see also Prada et al. 1994). To test our models, we have compared the synthetic profiles
 of the Balmer lines with the observations (see Gonz\'alez Delgado et al 1997 for 
details of the observations). The models are binned to the spectral resolution of the 
observations (the dispersion is 1.4 \AA/pix). The Balmer lines are partially filled 
with nebular emission; therefore the fitting has to be done based on the wings of the 
absorption features. This effect is less important for the higher terms of the Balmer 
series because the 
nebular emission drops quickly with decreasing wavelength. Figure 19 plots the observed
 lines and the synthetic models for a burst of 10 and 50 Myr at Z$\odot$/5 metallicity 
(assuming Salpeter IMF, $M_{low}$=1 M$\odot$ and $M_{up}$=80 M$\odot$). The profiles 
indicate that the Balmer lines are more compatible with a 10 Myr old burst than with one that is 50 Myr old. Ages older than 10 Myr produce profiles which are wider than the 
observed ones. This comparison confirms the result of Gonz\'alez Delgado et al. (1997)
 that the age of the SSC B is about 10 Myr and shows that this technique can discriminate well 
between a young and an intermediate-age population.  

\subsection{The nucleus of NGC 205}

We have observed the nucleus of the dwarf elliptical galaxy NGC 205 to provide a spectral template of an intermediate age population in order to analyze the stellar population of a sample of Seyfert 2 nuclei 
(Gonz\'alez Delgado et al 1998). The equivalent widths of the Balmer lines measured in 
our spectrum (Ew(H$\beta$)=7.7 \AA, Ew(H$\gamma$)=7.4 \AA, Ew(H$\delta$)=8.8 \AA, 
Ew(H$8$)=7.4 \AA) are weaker than the values predicted by a population in the range 100-500 Myr
(see Figure 14). Bica, Alloin, \& Schmidt (1990) have undertaken the population synthesis 
of the integrated optical light of the nucleus of NGC 205. They conclude that the 
dominant population is in the range 100-500 Myr; however, this population only produces 
$\simeq$ 50$\%$ of the optical light. The remaining 50$\%$ is due to young 
($\simeq$ 10 Myr), intermediate (1-5 Gyr) and old ($\geq$ 10 Gyr) components that 
contribute with 10$\%$, 20$\%$ and 20$\%$ of 
the optical light, respectively. These populations have equivalent widths of the Balmer 
lines that are weaker 
than the intermediate population of 100-500 Myr (at these ages, the Balmer lines have 
their maximum strength). Thus, the equivalent width of the Balmer lines is diluted by 
the contribution of the other components with respect to a single population of several
 hundred Myrs. We have combined our models contributing to the total light in the 
fraction derived by Bica et al (1990) and we found that the combined synthesis profile 
fits very well the observations (Figure 20).

\subsection{A luminous "E+A" galaxy}

Oegerle, Hill, \& Hoessel (1991) reported the serendipitous discovery of a relatively low redshift 
galaxy in the post-starburst phase that is one of the best examples of the 
so-called "E+A" galaxies (Dressler \& Gunn 1983). The spectra of these galaxies show very 
strong Balmer absorption lines produced by a large number of A stars. These stars result from a 
burst of star formation that probably occurred $\sim$1 Gyr ago, and are mixed with the underlying 
old stellar population typical of an elliptical galaxy.

The equivalent widths of the Balmer lines measured in the spectrum of the "E+A" galaxy discovered 
by Oegerle et al. (1991) are ew(H$\beta$)=9.5 \AA, ew(H$\gamma$)= 10.6 \AA, ew(H$\delta$)= 8.7 \AA, 
and ew(H8)= 8.6 \AA. These values are plotted as a horizontal dotted lines in Figure 18. With the 
exception of H$\delta$, the strength of the Balmer lines indicates that the galaxy is dominated by 
a population of A stars. The equivalent widths of the lines are well fitted by a burst that occurred in 
the galaxy 10$^8$ to 10$^9$ year ago. Profile shapes in spectra of higher resolution could discriminate between the 10$^8$ and 10$^9$ year age.

\section{Summary and conclusions}

We have computed evolutionary synthesis models that predict high-resolution spectra of a stellar 
popultation in the wavelength range: 3720-3920, 3990-4150, 4300-4400, 4420-4580 and 
4820-4950 \AA. These models predict the absorption line profiles of the hydrogen Balmer
series (H$\beta$, H$\gamma$, H$\delta$, H9, H10, H11, H12 and H13) and the neutral helium
lines  (HeI $\lambda$4922,  HeI $\lambda$4388, HeI $\lambda$4144, HeI $\lambda$4121,  HeI $\lambda$4026,  HeI $\lambda$4009 and HeI $\lambda$3819)  as a function of the age, IMF 
parameters and metallicity in two different star formation scenarios: an instantaneous burst and 
star formation proceeding continuously at a constant rate. 

Models are generated using the isochrone synthesis method assuming that stars evolve 
from the ZAMS following the set of evolutionary tracks of the Geneva group. At each age, the spectrum is computed assigning the corresponding 
high-resolution spectrum from our stellar library to each star. The library was
computed assuming that the atmosphere of the stars hotter than 25000 K is in NLTE and  
cooler stars in LTE and spans a range 
of T$_{eff}$ between 50000 to 4000 K and gravity 0.0 $\leq$ log g $\leq$ 5.0 (see paper I for details).

The computed profiles of the Balmer and HeI lines are useful for comparison with spectra of 
star-forming regions. High spectral resolution profiles of these lines (in the models and observations)
are required to constrain the evolutionary state of a very young stellar population through 
the analysis of the Balmer absorption wings under the nebular emission-lines.
Evolutionary synthesis profiles of the higher terms of the Balmer series are very useful,
as anticipated, to age-date the starburst because the nebular Balmer 
decrement decreases very rapidly with decreasing wavelength whereas the equivalent widths of 
the stellar Balmer absorption lines are almost constant with wavelength. A 4 Myr old burst shows a 
H8 stellar line of equal strength to the nebular emission-line; in contrast, at this age H$\beta$ 
is dominated by nebular emission. HeI $\lambda$3819 is also very useful to age-date starbursts 
because the nebular emission is weaker than in the other HeI lines and is dominated by the 
stellar absorption if the burst is older than 3 Myr. The profiles of the absorption features have 
been computed for the post-starburst phase up to 1 Gyr. These profiles can be used to date super star clusters and nuclei of galaxies dominated by an intermediate age population (younger than 1 Gyr). 

The strength of the Balmer absorption lines is sensitive to evolutionary effects. Their equivalent 
widths increase with age until about 
a few hundred Myr when the stellar population is dominated by early-A type stars; 
they range between 2 and 16 \AA\ if the cluster is formed at solar metallicity. 
During the first 4 Myr, the Balmer lines are not
sensitive to the age because the equivalent widths of the lines are almost constant, H$\beta$, 
H$\gamma$ and H$\delta$ $\simeq$ 3 \AA, and H9 and H10 $\simeq$ 2 \AA, respectively.
The Balmer lines are always weaker than 6 \AA\ if the starburst is in the nebular phase 
(younger than 10 Myr old). Balmer lines are also sensitive to the change of the IMF 
parameters during the first 4 Myr of the evolution of the starburst. Stronger 
lines are formed in the cluster if the mass of the cluster is distributed following 
a truncated Salpeter IMF with $M_{up}$=30 M$\odot$ or if the IMF is steeper than 
Salpeter. However, the strength of the lines is less sensitive to the low mass 
cut-off, contrary to previous suggestions. Balmer lines are prominent ($\simeq$
2 \AA) even when the low mass cut-off is as high as 10 M$\odot$. Weaker lines are formed
if the slope of the IMF is flatter than Salpeter because more massive stars are 
formed, which have weaker Balmer lines than early-type B stars. However, this effect 
is only important in the first few Myr of the evolution of the starburst when massive stars 
are still on the main sequence.

HeI lines are also sensitive to the evolution of the starburst. They increase (but 
not monotonically) with age until about 30-50 Myr; after this age, they decrease. 
No HeI lines are formed after 100 Myr because stars hotter than B8 have evolved from 
the main sequence, and stars cooler than B8 ($\simeq$ 12000 K) do not show HeI lines. 
The equivalent widths of HeI $\lambda$4471 and HeI $\lambda$4026, which are 
the strongest  HeI lines, range between 0.5 and 1.1 \AA. During the first 4 Myr 
of the evolution of the starburst, the strength of these lines is constant ($\simeq$
0.5 \AA\ for HeI $\lambda$4471 and HeI $\lambda$4026 and 0.3 \AA\ for
HeI $\lambda$4929, HeI $\lambda$4388 and HeI $\lambda$3819). We estimate that the correction 
to the HeI $\lambda$4471 nebular emission-lines due to the underlying absorption is 
between 5 and 25$\%$ if the nebular emission has an equivalent width between 10 and 2.5 \AA\ (a young burst).
The HeI lines are sensitive to the IMF parameters in the same way as the Balmer absorption lines.

For continuous star formation, the strength of the Balmer lines increases
with time until 500 Myr. The equivalent widths of the HeI lines increase with time until 100 Myr, then they turn over. The equivalent width of the stellar absorption lines ranges between 3 
and 10 \AA, and 0.5 and 0.9 \AA\
for the Balmer and HeI lines, respectively. The higher-order Balmer absorption lines are more 
useful to date-age starbursts than H$\beta$ because the latter is mainly dominated by the nebular emission; in contrast, H8 is dominated by the stellar absorption if continuous star formation lasts for more than 
30 Myr. HeI $\lambda$3819 is also very useful to date-age starbursts because the line is dominated 
by the stellar absorption if continuous star formation lasts for more than 20 Myr.

The advantages of using Balmer and HeI lines in absorption to date starbursts with respect to using nebular emission-lines are twofold: a) the age can be constrained in a much wider range, including the nebular and the post-nebular phases; and b) the strengths of the absorption lines are not affected by extinction or by the leaking of ionizing photons.
The models have been tested by comparing the strength of the Balmer lines with those
of the stellar clusters in the LMC, SSC B in NGC 1569, the nucleus of NGC 205 and a luminous 
"E+A" galaxy. Very good agreement is found. The full set of models is available at our website http://www.iaa.es/ae/e2.html and at http://www.stsci.edu/science/starburst/, 
or on request from the authors at rosa@iaa.es




{\bf Acknowledgments}

We thank Ivan Hubeny for his help in the initial phase of this project and for making 
his code available to the community, Enrique P\'erez, Angeles D\'\i az and Grazyna Stasi\'nska 
for their helpful suggestions 
and comments, and Bill Oegerle for sending us the spectrum of the "E+A" galaxy. RGD thanks 
the STScI for financial support during the preparation of part of this paper. This work was supported  the Spanish DGICYT grant PB93-0139, and by the NASA LTSA grant NAGW-3138.

\clearpage

\figcaption{Synthetic spectrum (solid line) from 3700 to 5000 \AA\ predicted for an instantaneous burst of 10$^6$ M$\odot$ formed following a Salpeter IMF between M$_{low}$= 1 M$\odot$ and M$_{up}$= 80 M$\odot$ at the age 20 Myr compared with the low resolution spectral energy distribution (dashed line) for the same model parameters.}

\figcaption{Synthetic spectra from 3700 to 5000 \AA\ predicted for an instantaneous 
burst formed following a Salpeter IMF between M$_{low}$= 1 M$\odot$ 
and M$_{up}$= 80 M$\odot$ at ages 3, 50, and 500 Myr. The horizontal lines indicate 
the windows used for measuring the equivalent widths.}

\figcaption{ Equivalent widths of the Balmer lines measured in a window of 60 \AA\  
width for H$\beta$, H$\gamma$ and H$\delta$ and 30 \AA\ width for H8, H9 and H10. 
The flux is integrated from the pseudo-continuum defined by fitting a first-order polynomial (except for the range 3720--3920 \AA) to the continuum windows defined in Table 3 of paper I.  }

\figcaption{Same as Figure 3 but for continuous star-formation models.}

\figcaption{ Equivalent widths of the HeI lines measured in a window of 14 \AA\  
width for HeI $\lambda$ 4921, 4471, 4388, of 11 \AA\ for HeI $\lambda$4026 and 7 \AA\ for HeI $\lambda$3819. The flux is integrated from the pseudo-continuum defined fitting a first-order polynomial (except for the range 3720--3920 \AA) to the continuum windows defined in Table 3 of paper I.  }

\figcaption{Isochrone at 14 Myr (a) and 25 Myr (b) for a burst of 10$^6$ M$\odot$ formed following a Salpeter IMF between M$_{low}$= 1 M$\odot$ and M$_{up}$= 80 M$\odot$ at solar metallicity. The size of the circle is scaled by the product of the number of stars in each mass (effective temperature) interval and the luminosity.}

\figcaption{Synthetic spectra in the spectral range 4015--4040 \AA\ normalized to the pseudo-continuum for stars with T$_{eff}$ equal to 20000 K, 10000 K and 6000 K. }

\figcaption{Profile of HeI $\lambda$4026 for an instantaneous burst formed following a Salpeter IMF between M$_{low}$= 1 M$\odot$ and M$_{up}$= 80 M$\odot$ at ages 4, 14 and 25 Myr. }

\figcaption{Same as Figure 5 but for continuous star-formation models (SFR= 1 M$\odot$ yr$^{-1}$).}

\figcaption{ Synthetic profiles of a 2 Myr old instantaneous burst formed with a mass distributed between $M_{low}$= 1 M$\odot$ and $M_{up}$= 80 M$\odot$
at solar metallicity following a power law IMF with index $\alpha$= 2.35 (full line), $\alpha$= 1.0 (dashed line) and $\alpha$= 3.3 (dotted line).}

\figcaption{Equivalent width of H$\beta$ for an instantaneous burst at solar metallicity formed following different assumptions about the IMF parameters. }

\figcaption{Same as  Figure 10 for the line HeI $\lambda$4471.}

\figcaption{Synthetic profile of a 2 Myr old instantaneous burst formed with a mass following a Salpeter 
IMF at solar metallicity with: $M_{low}$= 1 M$\odot$ and $M_{up}$= 80 M$\odot$ (continuous line), $M_{low}$= 5 M$\odot$ and $M_{up}$= 80 M$\odot$ (dotted line), and $M_{low}$= 1 M$\odot$ and $M_{up}$= 30 M$\odot$ (dashed line)}

\figcaption{ Equivalent width of the Balmer lines for an instantaneous burst at Z$\odot$ (continuous line) and Z= 0.001 (dashed line) metallicity formed following a standard Salpeter IMF.}

\figcaption{Same as Figure 13 but for HeI $\lambda$4471.}

\figcaption{Equivalent width of the Balmer lines (H$\beta$, H$\delta$ and H8) as a function of the age 
for an instantaneous burst (a) and continuous star formation (b). The nebular emission-lines (full line) 
are estimated from the number of Lyman continuum photons; the absorption stellar lines 
(dotted lines) are from the evolutionary synthesis models presented in this work. }

\figcaption{Equivalent width of the HeI lines ($\lambda$4471, $\lambda$4922, and $\lambda$3819) 
as a function of the age for an instantaneous burst (a) and continuous star formation (b). The nebular
emission-lines (full line) are estimated from the ratio HeI $\lambda$4471/H$\beta$ which is 
predicted using the photoionization code CLOUDY, and the absorption stellar lines (dotted lines) are 
from the evolutionary synthesis models. }

\figcaption{ Equivalent width of the Balmer lines a) H$\beta$, b) H$\gamma$, c) H$\delta$, d) H8 of an instantaneous burst formed following a standard Salpeter IMF at solar (continuous line) and Z= 0.001 (dashed line) metallicity. The equivalent widths of the templates Y1 ($\simeq$ 10 Myr old), Y2 ($\simeq$ 25 Myr old), Y3 ($\simeq$ 80 Myr old) and Y4 ($\simeq$ 200-500 Myr old) in the LMC are plotted as circles. Horithontal lines indicates the equivalent withs of the Balmer lines measured in the spectrum of the "E+A" galaxy reported by Oegerle et all (1991) }

\figcaption{ Normalized optical spectrum of the SSC B in NGC 1569 (continuous line) observed at the 4 m telescope in KPNO with a dispersion of 1.4 \AA/pix. The synthetic normalized spectra of an instantaneous burst 10 Myr (dashed line) and 50 Myr old (dotted line) formed following a standard Salpeter IMF at Z$\odot$/5 metallicity are plotted. Models are smoothed to the resolution of the observations.}

\figcaption{Normalized optical spectrum of the nucleus of NGC 205 (continuous line) observed at the 4 m telescope in KPNO with a dispersion of 1.4 \AA/pix. The composite synthetic normalized spectrum was obtained by adding instantaneous burst models that contribute by the following fractions: 10$\%$ of  a young ($\simeq$ 10 Myr old), 50$\%$ of an intermediate  ($\simeq$ 300 Myr old), 20$\%$ of a moderately old ($\simeq$ 1 Gyr old) and 20$\%$ of an old ($\geq$ 10 Gyr) population.}

\clearpage

\begin{deluxetable}{ccccccc}
\footnotesize
\tablecaption{Equivalent width of Balmer series (\AA) for an instantaneous burst at solar metallicity.
\tablenotemark{a}}
\tablewidth{0pt} 
\tablehead{ 
\colhead{age (Myr)} & \colhead{H$\beta$ (\AA)} & \colhead{H$\gamma$ (\AA)} &
 \colhead{H$\delta$ (\AA)} & \colhead{H8 (\AA)} & \colhead{H9 (\AA)} & 
\colhead{H10 (\AA)} \nl
} 
\startdata
     
0 &  2.7 &  2.8 &  3.2 &  2.5 &  2.0 &  1.7 \nl
1 &  2.6 &  2.8 &  3.1 &  2.4 &  2.0 &  1.7 \nl
2 &  2.6 &  2.7 &  3.1 &  2.3 &  2.0 &  1.7 \nl
3 &  2.7 &  2.7 &  3.2 &  2.4 &  2.1 &  1.9 \nl
4 &  3.4 &  3.2 &  4.0 &  3.1 &  2.8 &  2.5 \nl
5 &  4.9 &  4.2 &  5.4 &  4.3 &  4.6 &  3.0 \nl
6 &  4.6 &  3.8 &  5.0 &  3.8 &  3.7 &  3.0 \nl
7 &  5.8 &  3.8 &  5.9 &  4.3 &  4.1 &  3.3 \nl
8 &  4.0 &  4.3 &  5.7 &  4.6 &  4.3 &  3.6 \nl
9 &  4.5 &  4.4 &  6.1 &  4.7 &  4.4 &  3.7 \nl
10 &  4.5 &  4.2 &  6.0 &  4.6 &  4.2 &  3.7 \nl
11 &  4.5 &  4.1 &  6.1 &  4.7 &  4.2 &  3.8 \nl
12 &  4.6 &  4.5 &  6.2 &  4.8 &  4.2 &  3.8 \nl
13 &  5.5 &  4.8 &  6.7 &  5.1 &  4.6 &  4.0 \nl
14 &  5.8 &  5.6 &  7.2 &  5.6 &  5.7 &  4.2 \nl
15 &  6.8 &  6.8 &  8.1 &  6.7 &  6.5 &  4.9 \nl
16 &  5.7 &  5.7 &  6.7 &  5.5 &  5.1 &  4.1 \nl
17 &  5.6 &  5.7 &  6.6 &  5.3 &  4.9 &  4.1 \nl
18 &  5.7 &  5.7 &  6.7 &  5.4 &  4.9 &  4.1 \nl
19 &  5.9 &  6.0 &  7.0 &  5.6 &  5.2 &  4.3 \nl
20 &  6.1 &  6.1 &  7.1 &  5.8 &  5.3 &  4.4 \nl
25 &  6.7 &  6.7 &  7.8 &  6.4 &  6.0 &  4.8 \nl
30 &  7.2 &  7.4 &  8.7 &  6.9 &  6.4 &  5.2 \nl
35 &  7.5 &  7.8 &  9.1 &  7.2 &  6.6 &  5.4 \nl
40 &  7.5 &  7.9 &  9.2 &  7.2 &  6.7 &  5.3 \nl
45 &  7.7 &  8 &  9.4 &  7.3 &  6.8 &  5.4 \nl
50 &  7.6 &  7.9 &  9.3 &  7.1 &  6.7 &  5.3 \nl
60 &  7.6 &  7.9 &  9.4 &  7.1 &  6.6 &  5.3 \nl
70 &  7.9 &  8.2 &  9.7 &  7.4 &  6.9 &  5.6 \nl
80 &  8.1 &  8.4 &  10 &  7.6 &  7 &  5.6 \nl
90 &  8.5 &  8.8 &  10.4 &  7.9 &  7.2 &  5.8 \nl
100 &  8.7 &  9.0 &  10.6 &  8.1 &  7.4 &  5.9 \nl
200 &  10.5 &  10.9 &  13.0 &   &  9.1 &  6.9 \nl
300 &  11.5 &  12.4 &  14.8 &  11.1 &  10.1 &  7.5 \nl
400 &  12.3 &  13.0 &  15.8 &  11.7 &  10.5 &  7.5 \nl
500 &  11.6 &  13.0 &  16.2 &  11.5 &  10.6 &  7.3 \nl
600 &  11.1 &  12.4 &  15.8 &  11.2 &  10.4 &  7.1 \nl
700 &  10.4 &  11.1 &  14.5 &  10.5 &  10.2 &  6.8 \nl
800 &  9.7 &  10.2 &  14.5 &  10.0 &  10.0 &  6.6 \nl
900 &  8.8 &  8.8 &  12.1 &  9.3 &  9.8 &  6.2 \nl
1000 &  8.3 &  8.1 &  11.6 &  8.8 &  9.7 &  6.1 \nl
\enddata
\tablenotetext{a} {The mass of the cluster is distributed 
following a  Salpeter IMF between  a $M_{low}$= 1 M$\odot$ and $M_{up}$= 80 
M$\odot$.}
\end{deluxetable}

\begin{deluxetable}{ccccccc}
\footnotesize
\tablecaption{Equivalent width of the Balmer series (\AA) for continuous star formation at solar metallicity  \tablenotemark{a}}
\tablewidth{0pt}
\tablehead{
\colhead{age (Myr)} & \colhead{H$\beta$ (\AA)} & \colhead{H$\gamma$ (\AA)} 
& \colhead{H$\delta$ (\AA)} & \colhead{H8 (\AA)} & \colhead{H9 (\AA)} & 
\colhead{H10 (\AA} \nl
} 
\startdata

0 &  2.6 &  2.8 &  3.2 &  2.5 &  2.0 &  1.7 \nl
1 &  2.6 &  2.8 &  3.1 &  2.4 &  2.0 &  1.7 \nl
2 &  2.6 &  2.8 &  3.1 &  2.4 &  2.0 &  1.8 \nl
3 &  2.6 &  2.7 &  3.1 &  2.4 &  2.0 &  1.8 \nl
4 &  2.7 &  2.9 &  3.3 &  2.5 &  2.2 &  1.9 \nl
5 &  3.2 &  3.0 &  3.7 &  2.9 &  2.6 &  2.2 \nl
6 &  3.4 &  3.1 &  3.8 &  3.0 &  2.8 &  2.2 \nl
7 &  3.6 &  3.3 &  4.1 &  3.2 &  3.0 &  2.4 \nl
8 &  3.8 &  3.3 &  4.2 &  3.3 &  3.1 &  2.4 \nl
9 &  3.8 &  3.4 &  4.3 &  3.4 &  3.2 &  2.5 \nl
10 &  3.8 &  3.5 &  4.4 &  3.4 &  3.2 &  2.5 \nl
15 &  4.1 &  3.7 &  4.8 &  3.7 &  3.5 &  2.8 \nl
20 &  4.3 &  4.1 &  5.1 &  4.0 &  3.7 &  3.0 \nl
30 &  4.6 &  4.4 &  5.6 &  4.3 &  4.1 &  3.3 \nl
40 &  4.8 &  4.7 &  6.0 &  4.7 &  4.3 &  3.6 \nl
50 &  5.0 &  5.1 &  6.3 &  4.9 &  4.5 &  3.8 \nl
60 &  5.1 &  5.3 &  6.5 &  5.0 &  4.7 &  3.9 \nl
70 &  5.1 &  5.5 &  6.6 &  5.2 &  4.8 &  4.0 \nl
80 &  5.2 &  5.7 &  6.7 &  5.2 &  4.9 &  4.1 \nl
90 &  5.4 &  5.9 &  6.9 &  5.3 &  4.9 &  4.1 \nl
100 &  5.6  &  6.1 &  7.1 &  5.4 &  5.0 & 4.2 \nl
200 &  6.5  &  6.6 &  7.9  &  6.3 &  5.7 &  4.7 \nl
300 &  7.1  &  7.2 &  8.6  &  6.6  & 6.0  & 4.8 \nl
400 &  7.5  &  7.6 &  9.2  &  7.0 &  6.4 &  5.1\nl
500 &  7.8  &  8.0 &  9.6  &  7.2 &  6.6 &  5.2\nl
600 &  8.0  &  8.2 & 10.0 &  7.4 &  6.8 &  5.3\nl
700 &  8.2  &  8.3 & 10.2 &  7.5  &  6.9 &  5.3\nl
800 &  8.2  &  8.4 & 10.3 &  7.6 &  7.0 &  5.4\nl
900 &  8.2  &  8.5 & 10.3 &  7.6  &  7.0 &  5.4 \nl
1000& 8.2  &  8.4 & 10.3 &  7.6 &  7.1 &  5.3\nl
\enddata
\tablenotetext{a} {The mass is distributed 
following a Salpeter IMF between  a $M_{low}$= 1 M$\odot$ and $M_{up}$= 80 M$\odot$.}
\end{deluxetable}

\begin{deluxetable}{cccccc}
\footnotesize
\tablecaption{Equivalent width of HeI (\AA) for an instantaneous burst at solar metallicity}
\tablewidth{0pt}
\tablehead{
\colhead{age (Myr)} & \colhead{$\lambda$4922 (\AA)} & \colhead{$\lambda$4471 (\AA)} 
& \colhead{$\lambda$4388 (\AA)} & 
\colhead{$\lambda$4026 (\AA)} & \colhead{$\lambda$3819 (\AA)}  \nl
} 
\startdata

0 &  0.25 &  0.47 &  0.19 &  0.52 &  0.29 \nl
1 &  0.24 &  0.47 &  0.18 &  0.52 &  0.29 \nl
2 &  0.28 &  0.51 &  0.19 &  0.55 &  0.32 \nl
3 &  0.40 &  0.60 &  0.25 &  0.60 &  0.40 \nl
4 &  0.43 &  0.61 &  0.32 &  0.64 &  0.41 \nl
5 &  0.96 &  0.91 &  0.73 &  0.94 &  0.64 \nl
6 &  0.90 &  0.87 &  0.64 &  0.81 &  0.59 \nl
7 &          &  1.2 &  0.96 &  1.0 &  0.77 \nl
8 &  0.52 &  0.80 &  0.47 &  0.87 &  0.65 \nl
9 &  0.61 &  0.89 &  0.50 &  0.94 &  0.71 \nl
10 &  0.60 &  0.94 &  0.49 &  0.97 &  0.75 \nl
11 &          &  0.94 &  0.43 &  1.0 &  0.73 \nl
12 &  0.49 &  0.94 &  0.40 &  1.0 &  0.74 \nl
13 &          &  1.2 &  0.70 &  1.1 &  0.85 \nl
14 &  0.92 &  1.1 &  0.81 &  1.2 &  0.92 \nl
15 &  0.67 &  0.90 &  0.69 &  1.0 &  0.64 \nl
16 &  0.49 &  0.76 &  0.52 &  0.82 &  0.53 \nl
17 &  0.49 &  0.80 &  0.49 &  0.89 &  0.59 \nl
18 &  0.49 &  0.81 &  0.50 &  0.90 &  0.59 \nl
19 &  0.49 &  0.82 &  0.52 &  0.91 &  0.59 \nl
20 &  0.51 &  0.83 &  0.54 &  0.92 &  0.62 \nl
25 &  0.58 &  0.90 &  0.60 &  1.0 &  0.66 \nl
30 &  0.58 &  0.93 &  0.62 &  1.1 &  0.70 \nl
35 &  0.61 &  0.94 &  0.64 &  1.1 &  0.70 \nl
40   &  0.63 &  0.94 &  0.65 &  1.1 &  0.69 \nl
45   &  0.63 &  0.92 &  0.65 &  1.1 &  0.69 \nl
50   &  0.61 &  0.91 &  0.63 &  1.1 &  0.65 \nl
60   &          &  0.86 &  0.62 &  1.0 &  0.60 \nl
70   &  0.56 &  0.80 &  0.60 &  0.90 &  0.61 \nl
80   &         &  0.76 &  0.58 &  0.86 &  0.56 \nl
90   &         &  0.69 &  0.56 &  0.78 &  0.46 \nl
100 &   &  0.68 &  0.56 &  0.74 &  0.44 \nl
\enddata
\end{deluxetable}

\begin{deluxetable}{cccccc}
\footnotesize
\tablecaption{Equivalent width of HeI (\AA) for continuous star formation  at solar metallicity}
\tablewidth{0pt}
\tablehead{
\colhead{age (Myr)} & \colhead{$\lambda$4922 (\AA)} & \colhead{$\lambda$4471 (\AA)} 
& \colhead{$\lambda$4388 (\AA)} & 
\colhead{$\lambda$4026 (\AA)} & \colhead{$\lambda$3819 (\AA)}  \nl
}
\startdata

0    &  0.25 &  0.47 &  0.21 &  0.50 &  0.29 \nl
1.0 &  0.25 &  0.46 &  0.21 &  0.50 &  0.29 \nl
2.0 &  0.25 &  0.48 &  0.22 &  0.51 &  0.30 \nl
3.0 &  0.29 &  0.52 &  0.24 &  0.53 &  0.32 \nl
4.0 &  0.34 &  0.55 &  0.25 &  0.57 &  0.35 \nl
5.0 &  0.44 &  0.59 &  0.32 &  0.61 &  0.38 \nl
6.0 &  0.54 &  0.64 &  0.37 &  0.64 &  0.42 \nl
7.0 &  0.60 &  0.68 &  0.43 &  0.68 &  0.45 \nl
8.0 &  0.64 &  0.70 &  0.44 &  0.70 &  0.45 \nl
9.0 &  0.64 &  0.71 &  0.44 &  0.70 &  0.47 \nl
10 &  0.64 &  0.73 &  0.45 &  0.72 &  0.48 \nl
15 &  0.65 &  0.76 &  0.47 &  0.76 &  0.54 \nl
20 &  0.63 &  0.77 &  0.48 &  0.78 &  0.53 \nl
30 &  0.63 &  0.79 &  0.52 &  0.80 &  0.56 \nl
40 &  0.62 &  0.82 &  0.54 &  0.81 &  0.58 \nl
50 &  0.62 &  0.85 &  0.58 &  0.83 &  0.59 \nl
60 &  0.62 &  0.90 &  0.59 &  0.83 &  0.59 \nl
70 &  0.62 &  0.90 &  0.59 &  0.84 &  0.58 \nl
80 &  0.62 &  0.91 &  0.59 &  0.82 &  0.59 \nl
90 &  0.61 &  0.90 &  0.60 &  0.82 &  0.57 \nl
100& 0.62 &  0.89 &  0.61 &  0.83 &  0.57 \nl
200& 0.62 &  0.75 &  0.57 &  0.82 &  0.53 \nl
300& 0.61 &  0.71 &  0.56 &  0.77 &  0.53 \nl
400& 0.60 &  0.69 &  0.54 &  0.71 &  0.47 \nl
500& 0.60 &  0.67 &  0.55 &  0.71 &  0.49 \nl
600& 0.61 &  0.67 &  0.56 &  0.74 & 0.48 \nl
700& 0.62 &  0.67 &  0.59 &  0.72 & 0.50 \nl
800& 0.63 &  0.70 &  0.60 &  0.72 & 0.49 \nl
900& 0.63 &  0.70 &  0.60 &  0.72 & 0.50 \nl
1000&0.63&  0.70 &  0.60 &  0.72 & 0.50 \nl

\enddata
\end{deluxetable}

\begin{deluxetable}{ccccccc}
\footnotesize
\tablecaption{Equivalent width of the Balmer series (\AA) for an instantaneous burst at a metallicity Z= 0.001}
\tablewidth{0pt}
\tablehead{
\colhead{age (Myr)} & \colhead{H$\beta$ (\AA)} & \colhead{H$\gamma$ (\AA)} 
& \colhead{H$\delta$ (\AA)} & \colhead{H8 (\AA)} & \colhead{H9 (\AA)} & 
\colhead{H10 (\AA)} \nl
} 
\startdata

0 & 2.2 & 2.5 & 2.8 & 2.2 & 1.8 & 1.4 \nl
1 & 2.2 & 2.4 & 2.8 & 2.1 & 1.7 & 1.4 \nl
2 & 2.1 & 2.3 & 2.5 & 2.0 & 1.6 & 1.35 \nl
3 & 2.2 & 2.3 & 2.5 & 2.0 & 1.6 & 1.37 \nl
4 & 4.2 & 3.8 & 4.3 & 3.9 & 3.7 & 3.0 \nl
5 & 3.5 & 3.4 & 3.9 & 3.1 & 2.9 & 2.5 \nl
6 & 3.3 & 3.3 & 3.9 & 3.0 & 2.7 & 2.4 \nl
7 & 3.4 & 3.3 & 4.2 & 3.1 & 2.7 & 2.5 \nl
8 & 3.6 & 3.5 & 4.4 & 3.3 & 2.8 & 2.7 \nl
9 & 3.7 & 3.7 & 4.5 & 3.4 & 2.9 & 2.7 \nl
10 & 3.8 & 3.6 & 4.6 & 3.4 & 3.0 & 2.8 \nl
15 & 4.6 & 4.5 & 5.5 & 4.2 & 3.8 & 3.4 \nl
20 & 5.2 & 4.8 & 6.2 & 4.8 & 4.3 & 3.7 \nl
25 & 5.6 & 5.1 & 6.5 & 5.2 & 5.1 & 4.0 \nl
30 & 5.8 & 5.6 & 6.7 & 5.3 & 4.8 & 4.1 \nl
40 & 6.4 & 6.2 & 7.3 & 5.9 & 5.4 & 4.4 \nl
50 & 6.9 & 6.8 & 8 & 6.3 & 5.8 & 4.8 \nl
60 & 7.3 & 7.3 & 8.4 & 6.7 & 6.3 &5.0 \nl
70 & 7.5 & 7.5 & 8.7 & 7.0 & 6.5 & 5.2 \nl
80 & 7.8 & 7.9 & 9.2 & 7.4 & 6.9 & 5.5 \nl
90 & 8.0 & 8.1 & 9.5 & 7.6 & 7.1 & 5.7 \nl
100 & 8.3 & 8.5 & 9.9 & 7.8 & 7.3 & 5.8 \nl
200 & 9.5 & 10.0 & 11.8 & 9.1 & 8.5 & 6.6 \nl
300 & 9.6 & 10.0 & 12.3 & 9.2 & 8.7 & 6.5 \nl
400 & 9.7 & 10.2 & 12.6 & 9.3 & 9.0 & 6.4 \nl
500 & 9.7 & 10.4 & 12.9 & 9.6 & 9.2 & 6.7 \nl
600 & 9.9 & 10.5 & 13.3 & 9.7 & 9.5 & 6.8 \nl
700 & 10.2 & 11.0 & 13.7 & 10.1 & 9.8 & 7.0 \nl
800 & 10.4 & 11.3 & 14.2 & 10.2 & 10.0 & 7.1 \nl
900 & 10.6 & 11.5 & 14.5 & 10.5 & 10.3 & 7.4 \nl
1000 & 10.8 & 11.8 & 15.1 & 10.8 & 10.5 & 7.3\nl
\enddata
\end{deluxetable}

\begin{deluxetable}{ccccc}
\footnotesize
\tablecaption{Equivalent width of HeI (\AA) for an instantaneous burst at a metallicity Z= 0.001}
\tablewidth{0pt}
\tablehead{
\colhead{age (Myr)} & \colhead{$\lambda$4471 (\AA)} 
& \colhead{$\lambda$4388 (\AA)} & 
\colhead{$\lambda$4026 (\AA)} & \colhead{$\lambda$3819 (\AA)}  \nl
}
\startdata

0    & 0.43 &  0.18 &  0.47 &  0.25 \nl
1    &  0.42 &  0.17 &  0.46 &  0.25 \nl
2    &  0.38 &  0.16 &  0.43 &  0.22 \nl
3    &  0.43 &  0.17 &  0.44 &  0.24 \nl
4    &  0.36 &  0.35 &  0.38 &  0.19 \nl
5    &  0.60 &  0.36 &  0.64 &  0.40 \nl
6.   &  0.68 &  0.38 &  0.70 &  0.40 \nl
7    &  0.79 &  0.41 &  0.74 &  0.49 \nl
8    &  0.81 &  0.43 &  0.80 &  0.54 \nl
9    &  0.82 &  0.45 &  0.84 &  0.55 \nl
15 &   0.83 &  0.50 &  0.88 &  0.57 \nl
20 &   0.88 &  0.54 &  1.0 &  0.62 \nl
25 &   0.93 &  0.57 &  1.0 &  0.63 \nl
30 &   1.0   &  0.62 &  1.0 &  0.68 \nl
40 &   1.0   &  0.68 &  1.0 &  0.70 \nl
50 &   0.97 &  0.70 &  1.0 &  0.67 \nl
60 &   0.93 &  0.67 &  1.0 &  0.64 \nl
70 &   0.91 &  0.69 &  1.0 &  0.62 \nl
80 &   0.89 &  0.70 &  0.95 &  0.61 \nl
90 &   0.86 &  0.70 &  0.93 &  0.57 \nl
100 & 0.85 &  0.67 &  0.90 &  0.57 \nl
200 & 0.75 &  0.70 &  0.85 &  0.48 \nl
300 & 0.72 &  0.75 &  0.80 &  0.38 \nl
400 & 0.65 &  0.72 &  0.74 &  0.39 \nl
500 & 0.60 &  0.70 &  0.63 &  0.37 \nl
600 & 0.61 &  0.69 &  0.64 &  0.45 \nl
700 & 0.59 &  0.73 &  0.60 &  0.47 \nl
800 & 0.57 &  0.70 &  0.60 &  0.47 \nl
900 & 0.59 &  0.73 &  0.58 &  0.51 \nl
1000 &  0.59 &  0.81 &  0.57 &  0.52 \nl
\enddata
\end{deluxetable}

\begin{deluxetable}{cccccccccccc}
\footnotesize
\tablecaption{Equivalent width of the nebular emission of Balmer series (\AA) and HeI lines (\AA) for an instantaneous burst  at solar metallicity \tablenotemark{a}}
\tablewidth{0pt}
\tablehead{
\colhead{age} & \colhead{Q} & \colhead{L(H$\beta$)} & \colhead{HeI 4471/H$\beta$} & 
\colhead{H$\beta$} & \colhead{H$\gamma$}  & \colhead{H$\delta$} & \colhead{H8}  & \colhead{4922}   & \colhead{4471} 
 & \colhead{4026}  & \colhead{3819} \nl
\colhead{(1)} & \colhead{(2)} & \colhead{(3)} & \colhead{(4)} & 
\colhead{(5)} & \colhead{(6)}  & \colhead{(7)} & \colhead{(8)}  & \colhead{(9)}   & \colhead{(10)} 
 & \colhead{(11)}  & \colhead{(12)} \nl
} 
\startdata

0 &  52.7 &  40.4 &  0.048 &  400 &  134 &  64    &  22   &  5.3  &  14.7 &  5.3 &  2.6 \nl
1 &  52.7 &  40.4 &  0.046 &  367 &  124 &  58    &  20   &  4.8  &  13.1 &  4.6 &  2.3 \nl
2 &  52.7 &  40.4 &  0.037 &  308 &  102 &  48    &  16   &  3.2  &  8.8   &  3.1 &  1.5 \nl
3 &  52.5 &  40.2 &  0.028 &  120 &  39   &  18    &  6     & 0.9   &  2.5   &  0.9 &  0.4 \nl
4 &  52.2 &  39.9 &  0.046 &   68  &  23   &  11    &  3.6  & 0.9   &  2.4   &  0.8 &  0.4 \nl
5 &  51.9 &  39.6 &  0.050  &  40  &  16   &  8      &  2.8  &  0.5  &  1.7   &  0.7 &  0.35 \nl 
6 &  51.4 &  39.1 &  0.002  &  13  &   5    &  2.5   &  0.9  &  0.006 &0.02 &  0.007 &  0.004\nl
7 &  51.0 &  38.7 &  0         &  8    &  3.7  &  1.8   &  0.7  &  0      &  0 &  0 &  0 \nl
8 &  50.7 &  38.4 &  0         &  5    &  2.0  &  1.0   &  0.4  &  0      &  0 &  0 &  0 \nl
9 &  50.5 &  38.1 &  0         &  3.6 &  1.3  &  0.6   &  0.2  &  0       &  0 &  0 &  0 \nl
10 &  50.2 &  37.9 &  0       &  2.3 &  0.9  &  0.4   &  0.15 &  0      &  0 &  0 &  0 \nl
15 &  49.4 &  37.1 &  0       &  0.3 &  0.1  &  0.06 &  0.02 &  0       &  0 &  0 &  0 \nl
20 &  48.9 &  36.6 &  0       &  0.1 &  0.05 &  0.02 &  0.008&  0     &  0 &  0 &  0 \nl
\enddata
\tablenotetext{a} {(1): age (Myr); (2): Logarithm of the number of Lyman contiuum photons (s$^{-1}$); (3): Logarithm of the H$\beta$ luminosity (erg s$^{-1}$); (4): Line ratio of the nebular emission-lines HeI$\lambda$4471/H$\beta$ (4)-(8): Equivalent width of the Balmer lines (H$\beta$, H$\gamma$, H$\delta$ and H8, respectively); (9)-(12): Equivalent width of the nebular emission HeI lines ($\lambda$4922, $\lambda$4471, $\lambda$4026 $\lambda$3819) }
\end{deluxetable}

\begin{deluxetable}{cccccccccccc}
\footnotesize
\tablecaption{As Table 7 for continous star formation}
\tablewidth{0pt}
\tablehead{
\colhead{age} & \colhead{Q} & \colhead{L(H$\beta$)} & \colhead{HeI 4471/H$\beta$} & 
\colhead{H$\beta$} & \colhead{H$\gamma$}  & \colhead{H$\delta$} & \colhead{H8}  & \colhead{4922}   & \colhead{4471} 
 & \colhead{4026}  & \colhead{3819} \nl
\colhead{(1)} & \colhead{(2)} & \colhead{(3)} & \colhead{(4)} & 
\colhead{(5)} & \colhead{(6)}  & \colhead{(7)} & \colhead{(8)}  & \colhead{(9)}   & \colhead{(10)} 
 & \colhead{(11)}  & \colhead{(12)} \nl
} 
\startdata

1 &  52.7 &  40.4 &  0.047  &  379 &  127 &  60    &  21    &  4.9 &  14 &  4.9 &  2.4 \nl
2 &  53.0 &  40.7 &  0.046  &  356 &  120 &  56    &  19    &  4.5 &  13 &  4.5 &  2.2 \nl
3 &  53.2 &  40.8 &  0.044  &  295 &  98   &  46    &  16    &  3.5 &  9.9 &  3.5 &  1.7 \nl
4 &  53.2 &  40.9 &  0.045  &  227 &  75   &  35    &  12    &  2.8 &  7.8 &  2.7 &  1.3 \nl
5 &  53.3 &  40.9 &  0.045  &  182 &  61   &  29    &  10    &  2.2 &  6.3 &  2.3 &  1.1 \nl
6 &  53.3 &  41.0 &  0.045  &  155 &  54   &  26    &  9      &  1.9 &  5.5 &  2.0 &  1.00 \nl
7 &  53.3 &  41.0 &  0.045  &  135 &  48   &  23    &  8      &  1.7 &  4.9 &  1.8 &  0.90 \nl
8 &  53.3 &  41.0 &  0.045  &  125 &  45   &  21    &  7.5   &  1.5 &  4.6 &  1.7 &  0.85 \nl
9 &  53.3 &  41.0 &  0.045  &  119 &  43   &  20    &  7.1   &  1.5 &  4.4 &  1.6 &  0.80 \nl
10 &  53.3 &  41.0 & 0.045 &  114 &  41   &  20    &  6.9   &  1.4 &  4.2 &  1.5 &  0.78 \nl
15 &  53.3 &  41.0 & 0.045 &  96   &  35   &  17    &  5.9   &  1.2 &  3.6 &  1.3 &  0.66 \nl
20 &  53.3 &  41.0 & 0.045 &  83   &  30   &  14    &  5.0  &  1.0 &  3.1 &  1.1 &  0.57 \nl
30&  53.3 &  41.0 &  0.045 &  70   &  25   &  12    &  4.2  &  0.86 &  2.6 &  0.94 &  0.48 \nl
40 &  53.3 &  41.0 & 0.045 &  62   &  23   &  11    &  3.8  &  0.77 &  2.3 &  0.84 &  0.43 \nl
50 &  53.3 &  41.0 & 0.045 &  57   &  21   &  10    &  3.5  &  0.71 &  2.1 &  0.77 &  0.40 \nl
60 &  53.3 &  41.0 & 0.045 &  54   &  19   &  9      &  3.3  &  0.66 &  2.0 &  0.73 &  0.38 \nl
70 &  53.3 &  41.0 & 0.045 &  51   &  18   &  8.7   &  3.2  &  0.63 &  1.9 &  0.69 &  0.36 \nl
80 &  53.3 &  41.0 & 0.045 &  49   &  18   &  8.4   &  3.0  &  0.60 &  1.8 &  0.66 &  0.35 \nl
90&  53.3 &  41.0 & 0.045  &  47   &  17   &  8.0   &  2.9  &  0.58 &  1.7 &  0.63 &  0.33 \nl
100& 53.3 & 41.0 & 0.045  &  45   &  16   &  7.8   &  2.8  &  0.56 &  1.6 &  0.61 &  0.32 \nl
\enddata
\end{deluxetable}

%
%


\clearpage

\begin{thebibliography}{}

\bibitem[]{} Auer, L. H., \& Mihalas, D. 1972, ApJS, 205, 24
\bibitem[]{} Augarde, R., \& Lequeux, J. 1985, A\&A, 147, 273
\bibitem[]{} -- 1986a, A\&A, 162, 21 
\bibitem[]{} Bica, E., \& Alloin, D. 1986b, A\&AS, 66, 171
\bibitem[]{} Bica, E., Alloin, D., \& Schmidt, A. 1990, A\&A, 228, 23
\bibitem[]{} Brodie, J.P., Schroder, L.L., Huchra, J. P., Phillips, A.C., Kissler-Patig, M., \& Forbes, D. 1998, AJ, 116, 691
\bibitem[]{} Cananzi, K., Augarde, R., \& Lequeux, J. 1993, A\&AS, 101, 599
\bibitem[]{} -. 1994, Spac. Sc. Rev., 66, 75 
\bibitem[]{} Charbonnel, C., Meynet, G., Maeder, A., Schaller, G., \& Schaerer, D. 1993, A\&AS, 101, 415.
\bibitem[]{} Cid Fernandes, R., Storchi-Bergmann, T., \& Schmitt, H. 1998, MNRAS, 297, 579
\bibitem[]{} D\'\i az, A. I. 1988, MNRAS, 231, 57
\bibitem[]{} Dressler, A., \& Gunn, J.E. 1983, 270, 7 
\bibitem[]{} Ferland, G.J. 1997, Hazy, a Brief Introduction to CLOUDY, 
University of  Kentucky,
 Department of Physiscs and Astronomy Internal Report 
\bibitem[]{} Garc\'{\i}a-Vargas, M.L., Gonz\'alez-Delgado, R.M., P\'erez, E.,
Alloin, D.,  D\'{\i}az, A.I., \& Terlevich, E. 1997, ApJ, 478, 112 
 \bibitem[]{} Garc\'{\i}a-Vargas, M.L., Bressan, A., \& D\'\i az, A. 1995, A\&AS, 112, 35 
\bibitem[]{} Gonz\'alez Delgado, R.M., Garc\'{\i}a-Vargas, M.L., Goldader, J., Leitherer, C., \& Pasquali, A. 1999, ApJ, 513, 707
\bibitem[]{} Gonz\'alez Delgado, R.M., Heckman, T., Leitherer, C., Meurer, G., 
Krolik, J, Wilson, A.S., Kinney, A.L., \& Koratkar, A. 1998, ApJ, 505, 174
\bibitem[]{} Gonz\'alez Delgado, R.M., Leitherer, C., Heckman, T., \& Cervi\~no, M. 1997, ApJ, 483, 705
\bibitem[]{} Gonz\'alez-Delgado, R.M., P\'erez, E., D\'{\i}az, A.I.,
Garc\'{\i}a-Vargas, M.L., Terlevich, E., \& V\'{\i}lchez, J.M. 1995, ApJ, 439, 604
\bibitem[]{} Ho, L. C., \& Filippenko, A. V. 1996, ApJ, 466, L83
\bibitem[]{} Hubeny, I. 1988, Compt. Phys. Comm., 52, 103
\bibitem[]{} Hubeny, I., \& Lanz, T. 1995a, ApJ, 439, 875
\bibitem[]{} -. 1995b, TLUSTY -A User's Guide
\bibitem[]{} Hubeny, I.,  Lanz, T., \& Jeffery, C.S. 1995, SYNSPEC -A User's Guide
\bibitem[]{} Kurucz, R.L. 1979, ApJS, 40,1
\bibitem[]{} --. 1993, CD-ROM 13, ATLAS9 Stellar Atmosphere Programs and 2 km/s Grid (Cambridge: Smithsonian Astrophys. Obs.)
 \bibitem[]{} Izotov, Y. I., Thuan, T. X., \& Lipovetsky, V. A. 1997, ApJS, 108, 1
\bibitem[]{} Leitherer, C., Schaerer, D., Goldader, J. D., Gonz\'alez Delgado, R.M., Robert, C., Foo Kune, D., de Mello, D. F., Devost, D., Heckman, T. M. 1999, ApJS, in press
\bibitem[]{} Leitherer, C.,  et al. 1996, PASP, 108, 1996
\bibitem[]{} Lejeune, T., Buser, R., \& Cuisinier, F. 1997, A\&AS, 125, 229
\bibitem[]{} Lennon, D. J., Dufton, P.L., \& Fitzsimmons, A. 1993, A\&AS, 97, 559
\bibitem[]{} McCall, M. L., Rybski, P. M., \& Shields, G.A. 1985, ApJS, 57, 1
\bibitem[]{} Meurer, G., Heckman, T., Leitherer, C., Kinney, A., Robert, C., \& Garnett, D. 1995, AJ, 110, 2665
\bibitem[]{} Meynet, G., Maeder, A., Schaller, G., Schaerer, D., \& Charbonnel, C. 1994, A\&AS, 103, 97
\bibitem[]{} O'Connell, R., Gallagher, J., \& Hunter, D. 1994, ApJ, 433, 65
\bibitem[]{} Oegerle, W. R., Hill, J. M., \& Hoessel, J. G., 1991, ApJ, 381, L9
\bibitem[]{} Olofsson, K. 1995, A\&AS, 111, 57
\bibitem{} Osterbrock, D.E., 1989, {\em Astrophysics of Gaseous Nebulae and Active Galactic Nuclei}, University science Press
\bibitem[]{} Prada, F., Greve, A., \& McKeith, C. 1994, A\&A, 288, 396
\bibitem[]{} Rieke, G., Lebofsky, M., Thompson, R., Low, F., \& Tokunaga, A. 1980, ApJ, 238, 24
\bibitem[]{} Satyapal, S., et al 1995, ApJ, 448, 611
\bibitem[]{} Satyapal, S., et al 1997, ApJ, 483, 148
\bibitem[]{} Scalo, J.M. 1990, in {\it Windows on Galaxies}, ed. A. Renzini, G. Fabbiano, \& J.S. Gallagher (Dordrecht: Kluwer) 125
\bibitem[]{} Schaerer, D., Charbonnel, C., Meynet, G., Maeder, A., \& Schaller, G. 1993a, A\&AS, 102, 339
\bibitem[]{} Schaerer, D., Meynet, G., Maeder, A., \& Schaller, G. 1993b, A\&AS, 98, 523
\bibitem[]{} Schaller, G., Schaerer, D., Meynet, G., \& Maeder, A. 1992, A\&AS, 96, 269
\bibitem[]{} Schmutz, W., Leitherer, C., \& Gruenwald, R. 1992, PASP, 104, 1164
\bibitem[]{} Shields, G. A., \& Searle, L. 1978, ApJ, 222, 821
\bibitem[]{} Shields, J.C., \& Filippenko, A.V. 1990, AJ, 100, 1034
\bibitem[]{} Storchi-Bergmann, T., Kinney, A., \& Challis, P. 1995, ApJ, 98, 103
\bibitem[]{} Terlevich, E., D\'\i az, A.I., Terlevich, R., Gonz\'alez Delgado, R.M., P\'erez, E., \& Garc\'\i a-Vargas, M.L. 1996, MNRAS, 279, 1219
\bibitem[]{} Viallefond, F., \& Thuan, T.X. 1983, ApJ, 269, 444
\bibitem[]{} Walborn, N. R., \& Fitzpatrick, E. L. 1990, PASP, 102, 379
\end{thebibliography}
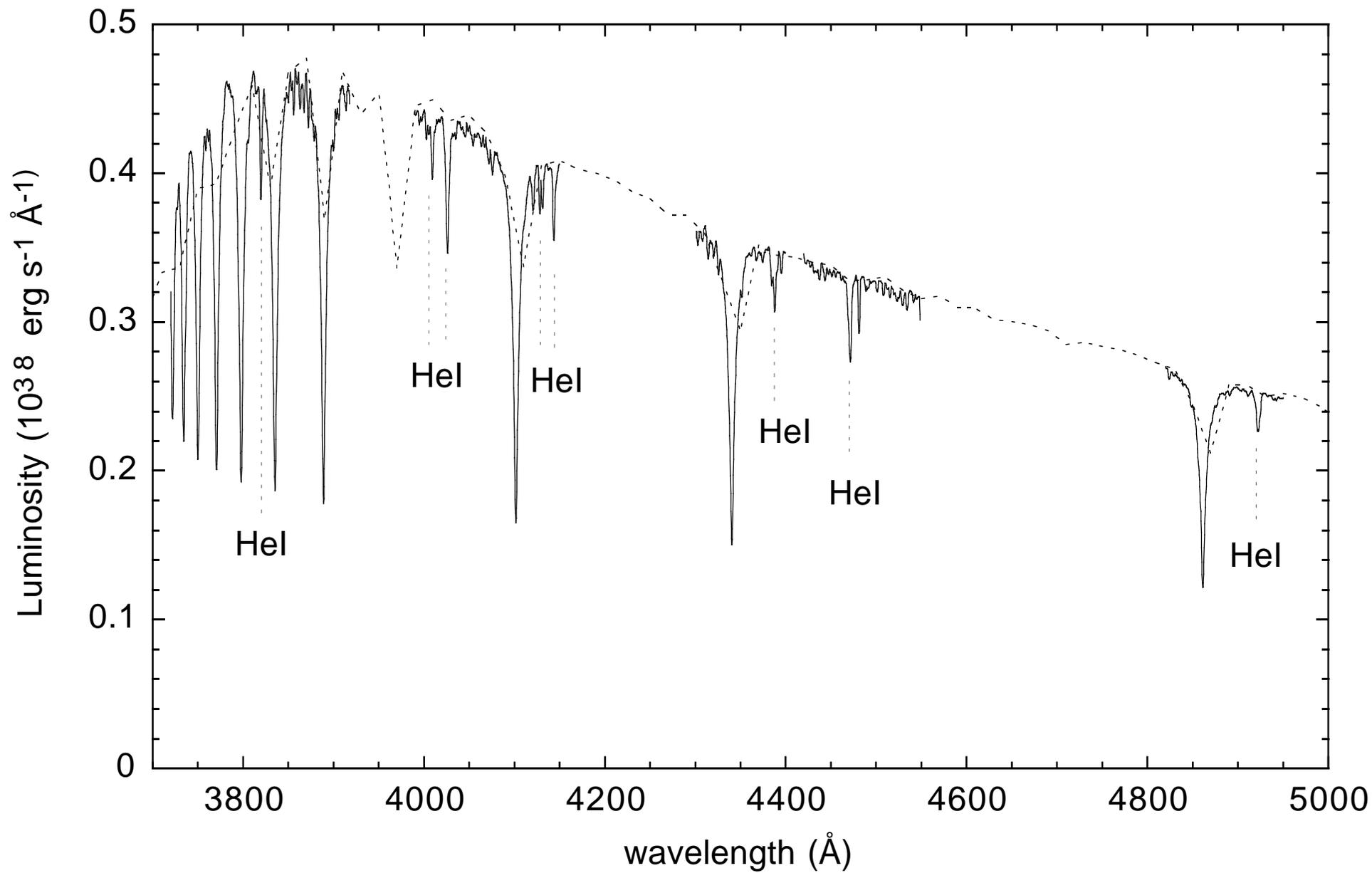
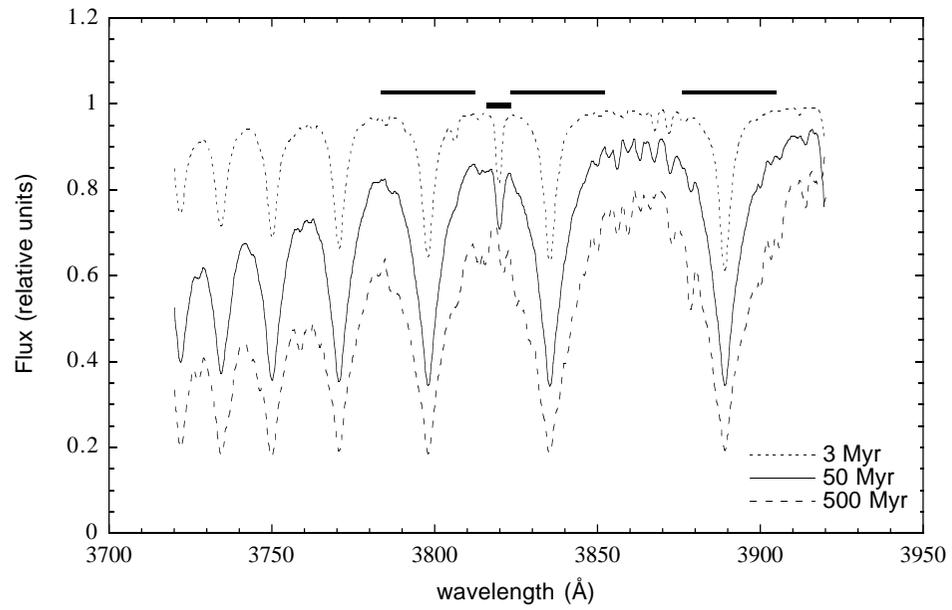
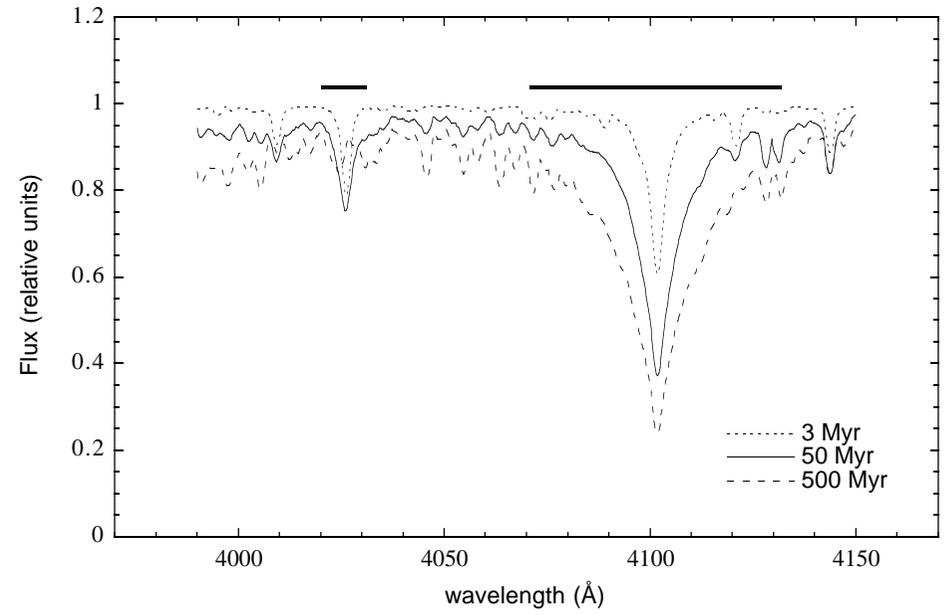
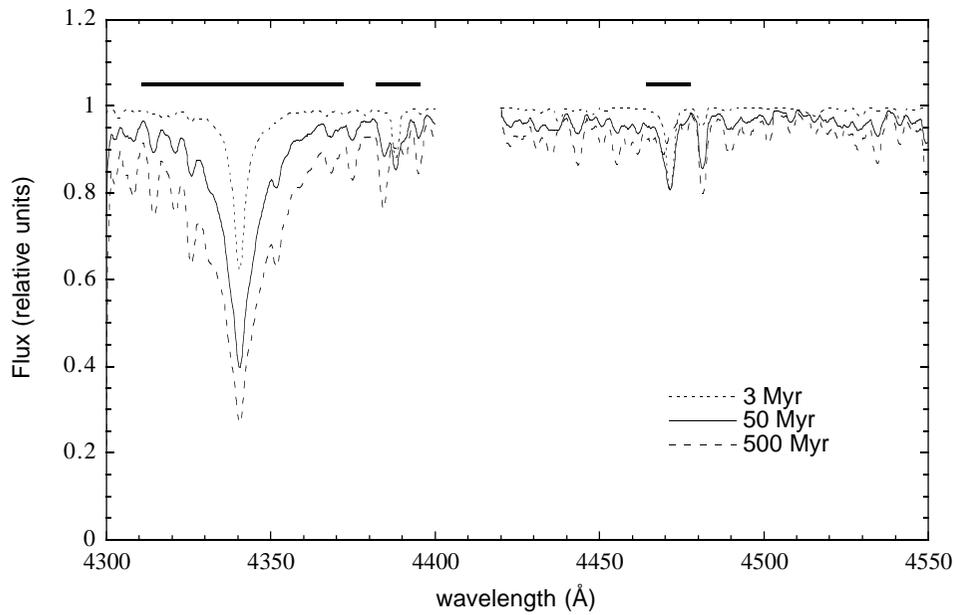
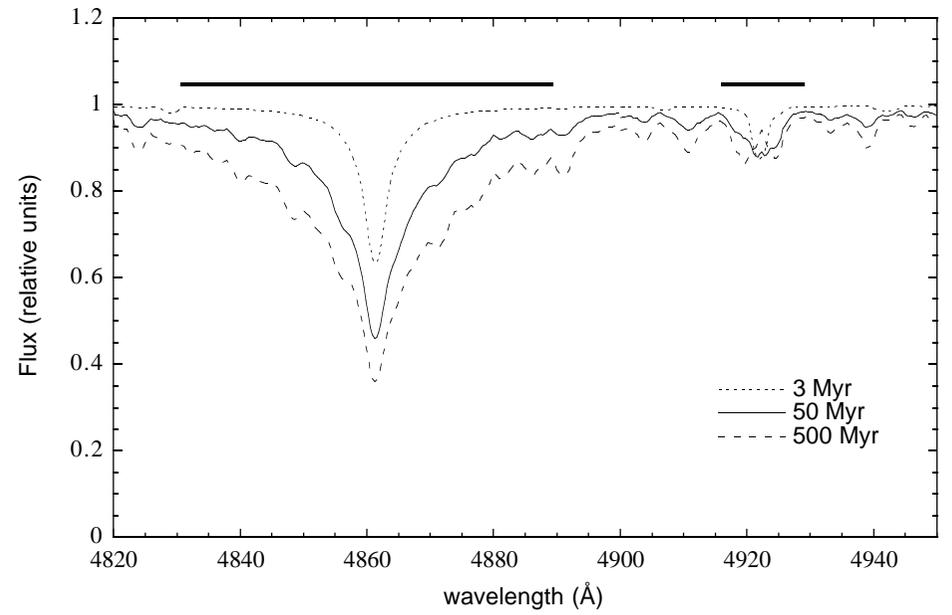
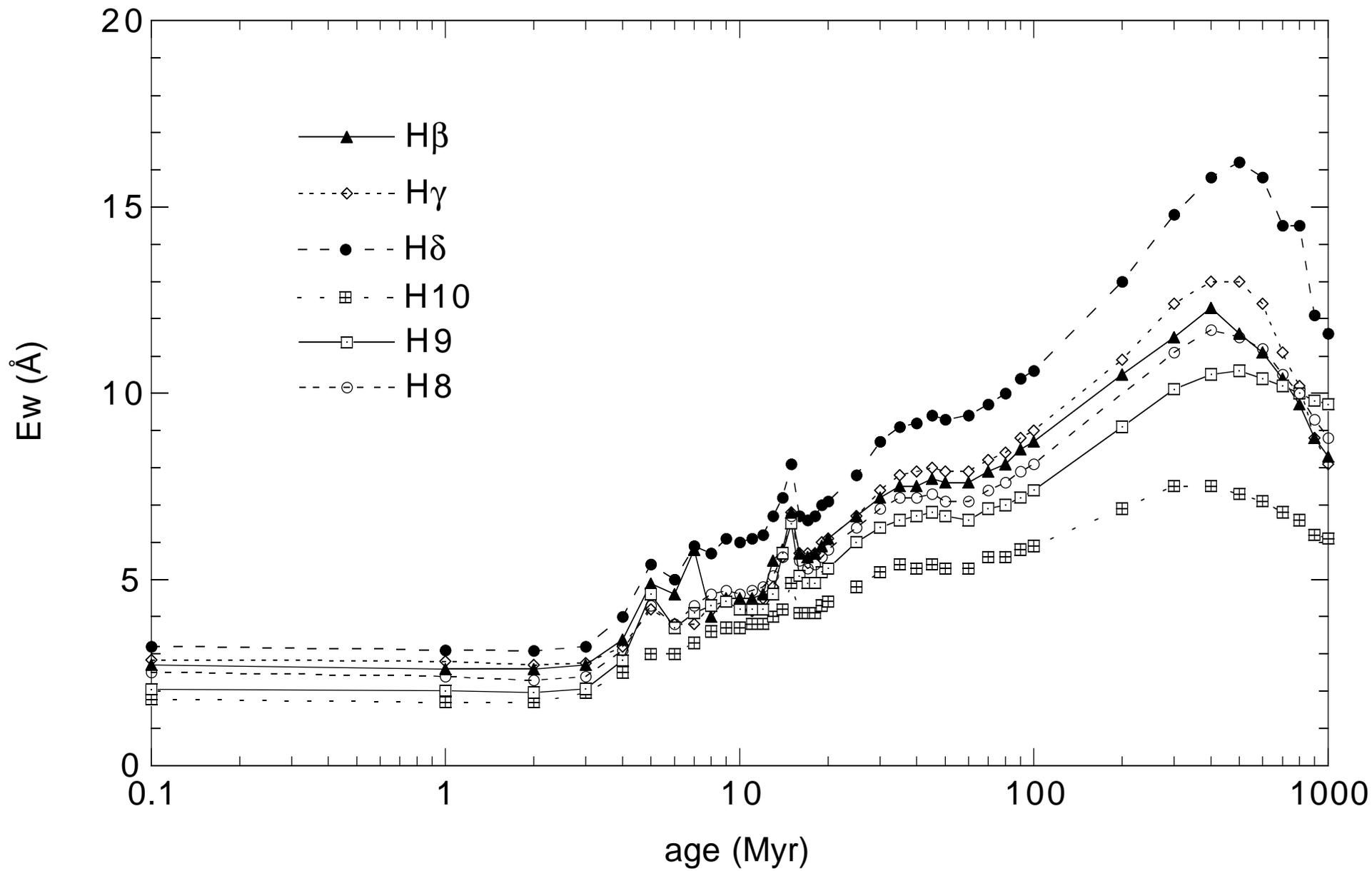
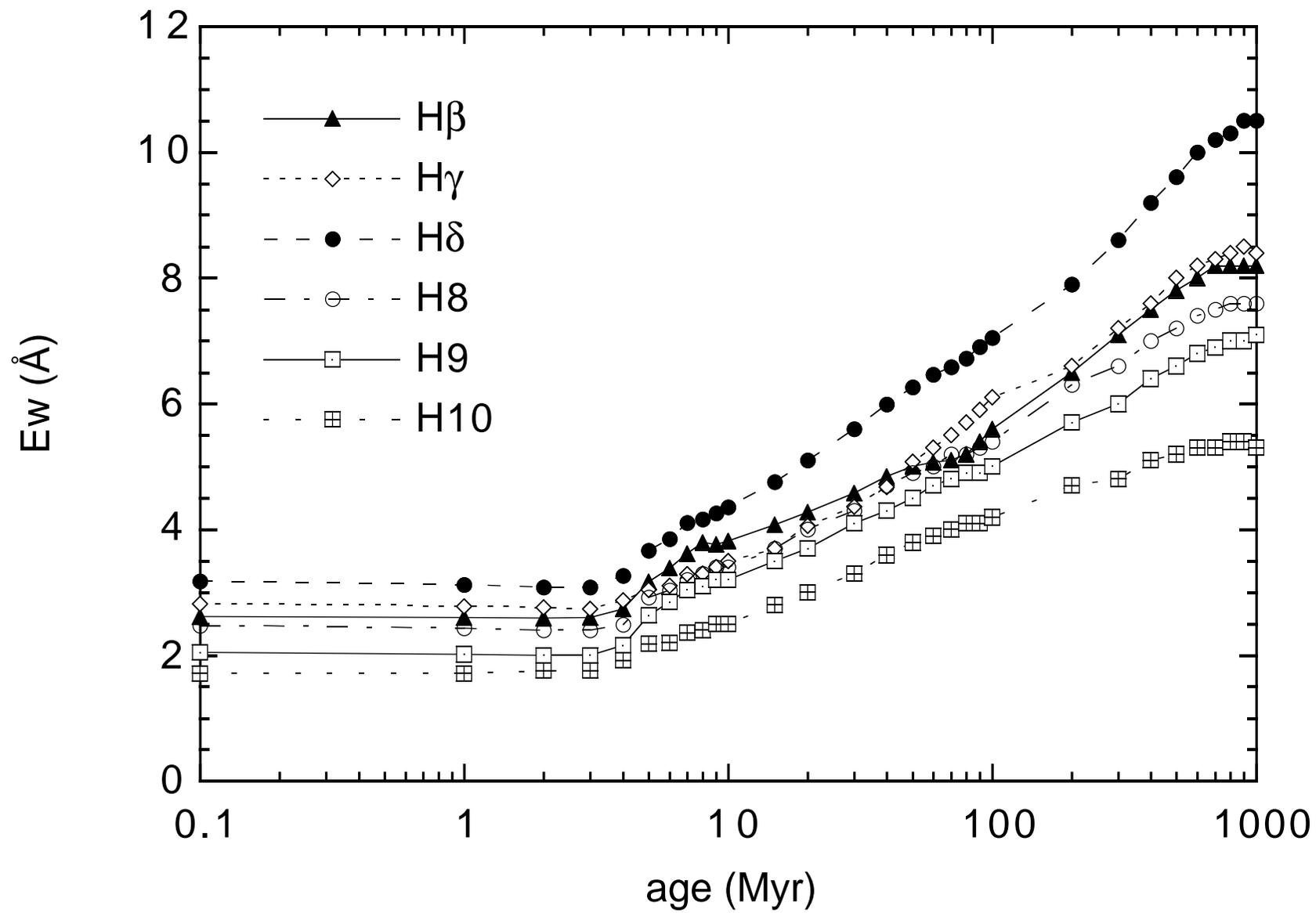
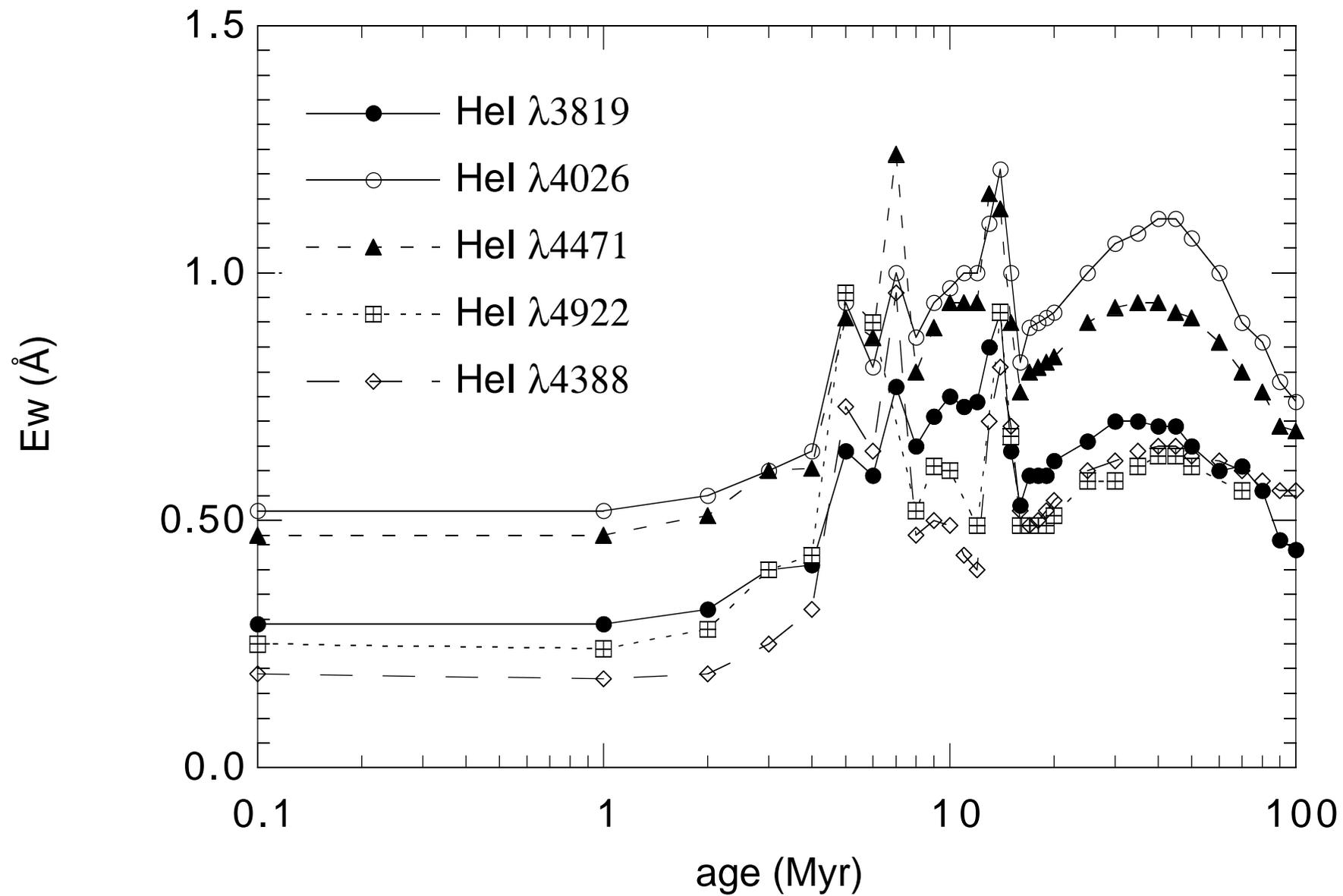
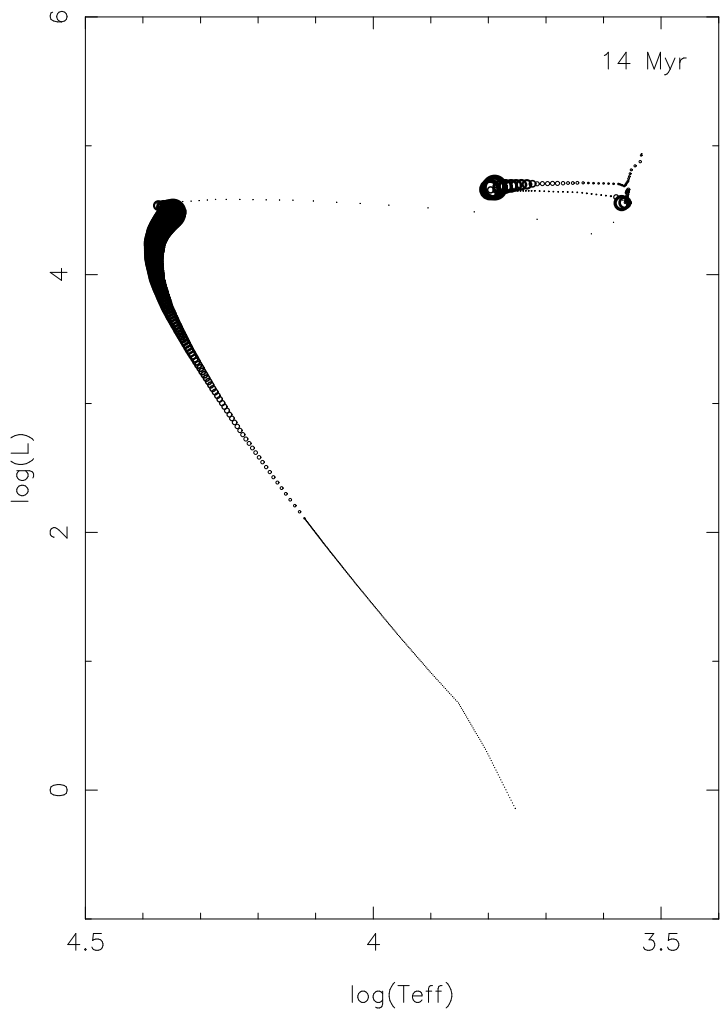
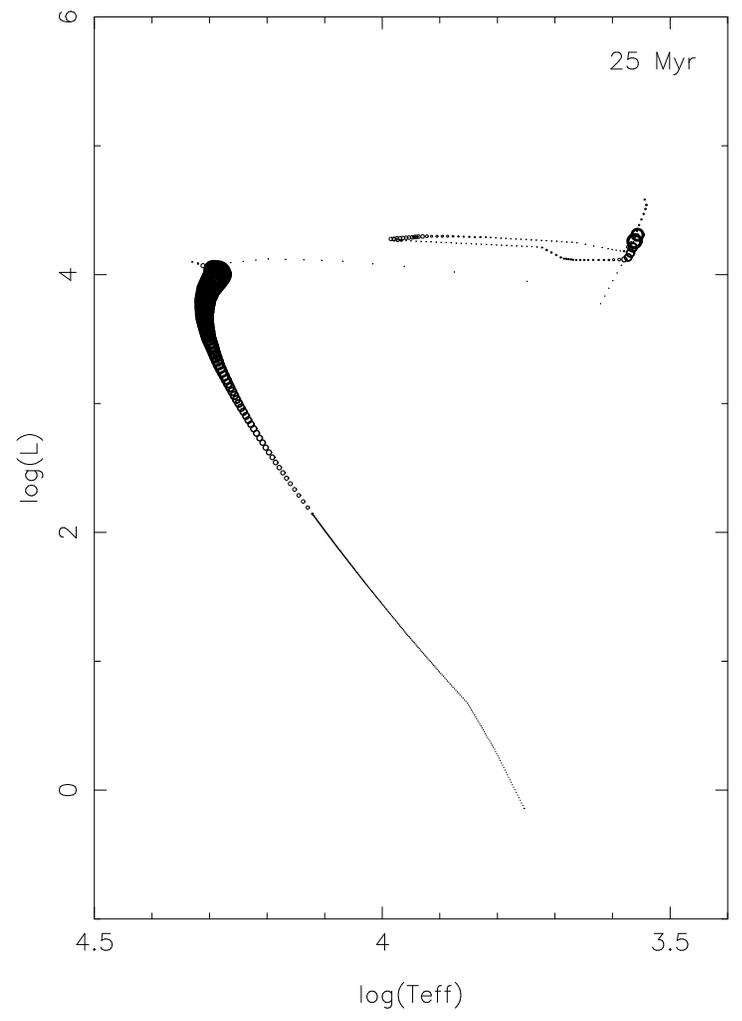
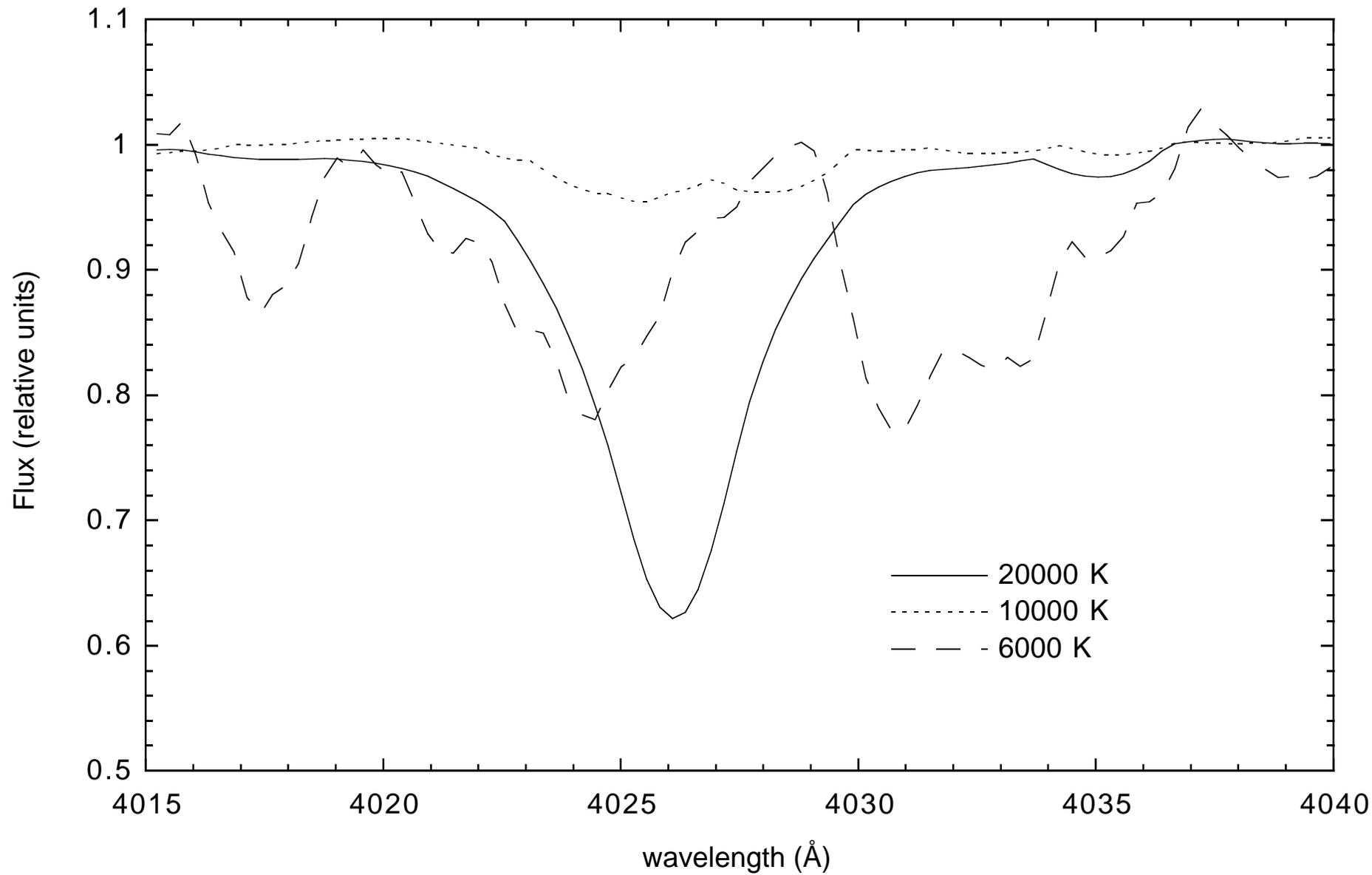
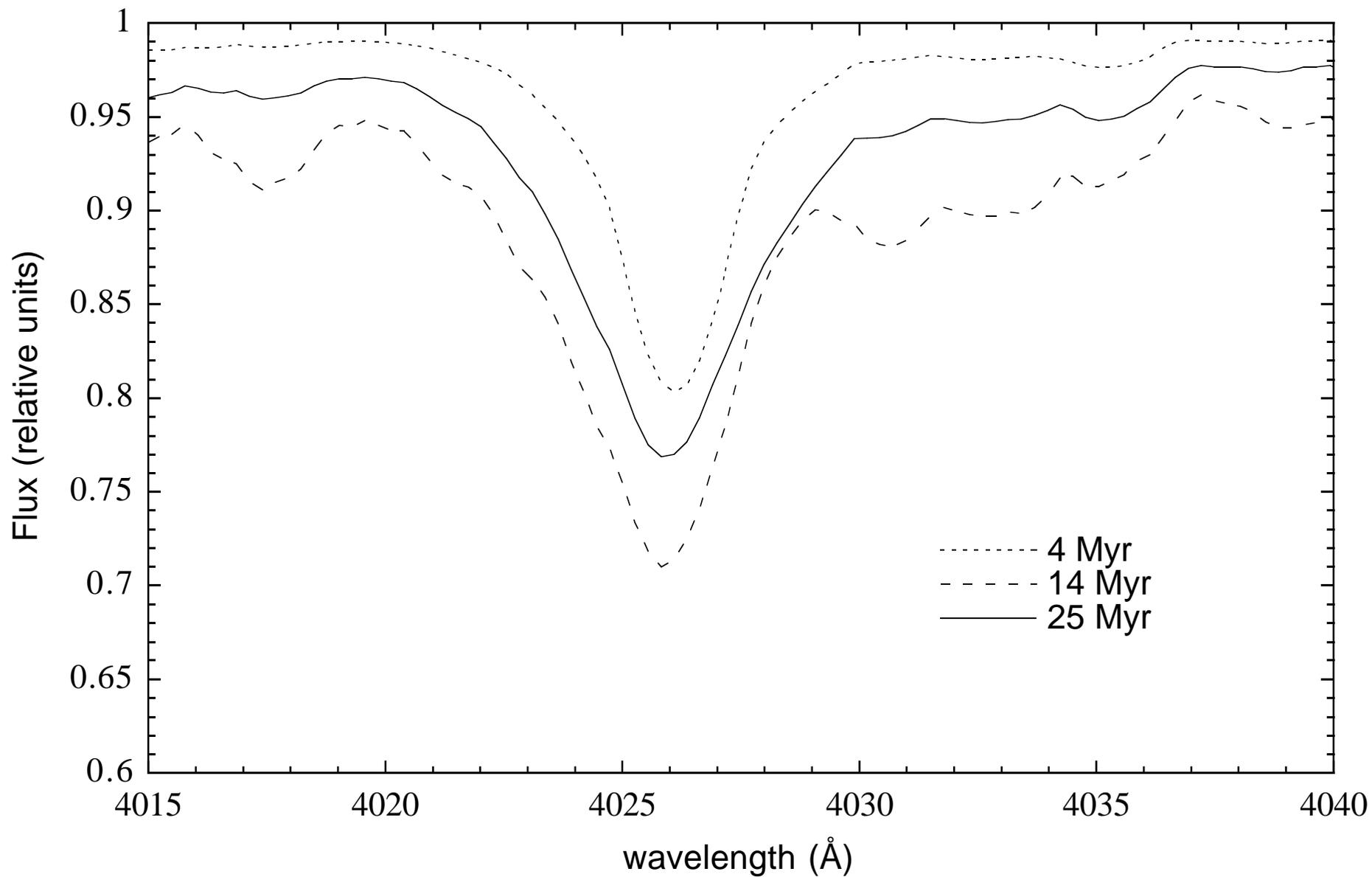
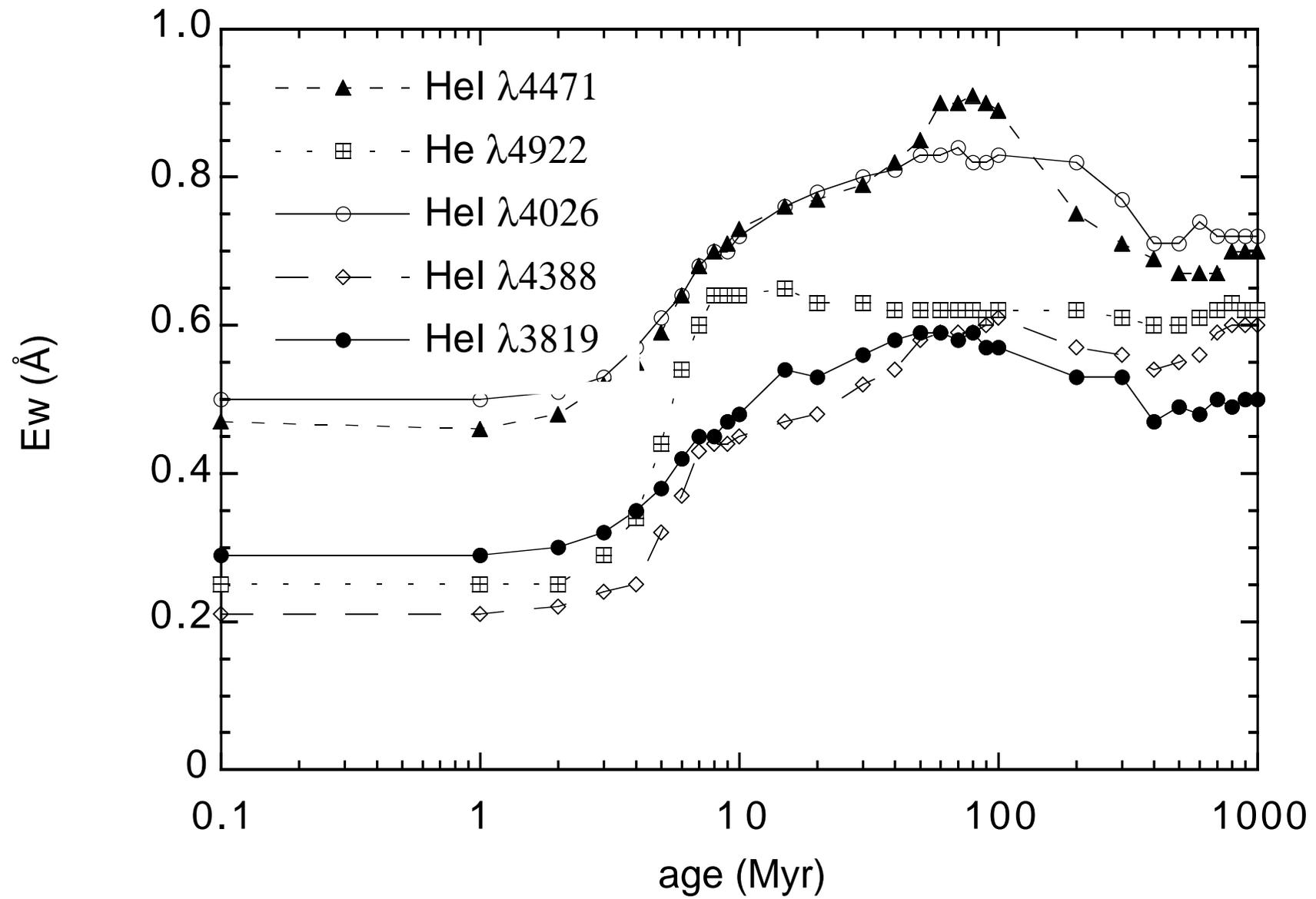
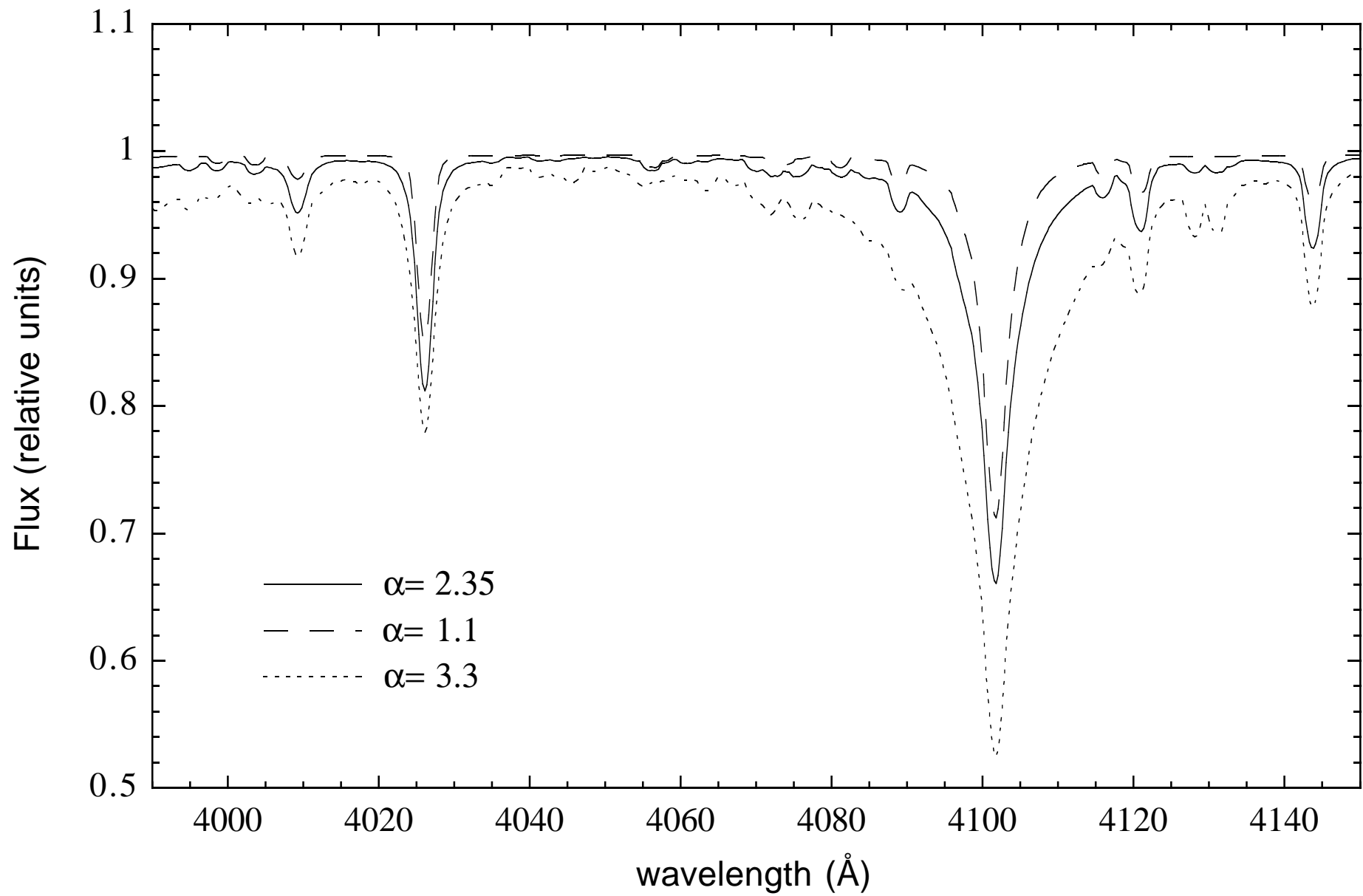
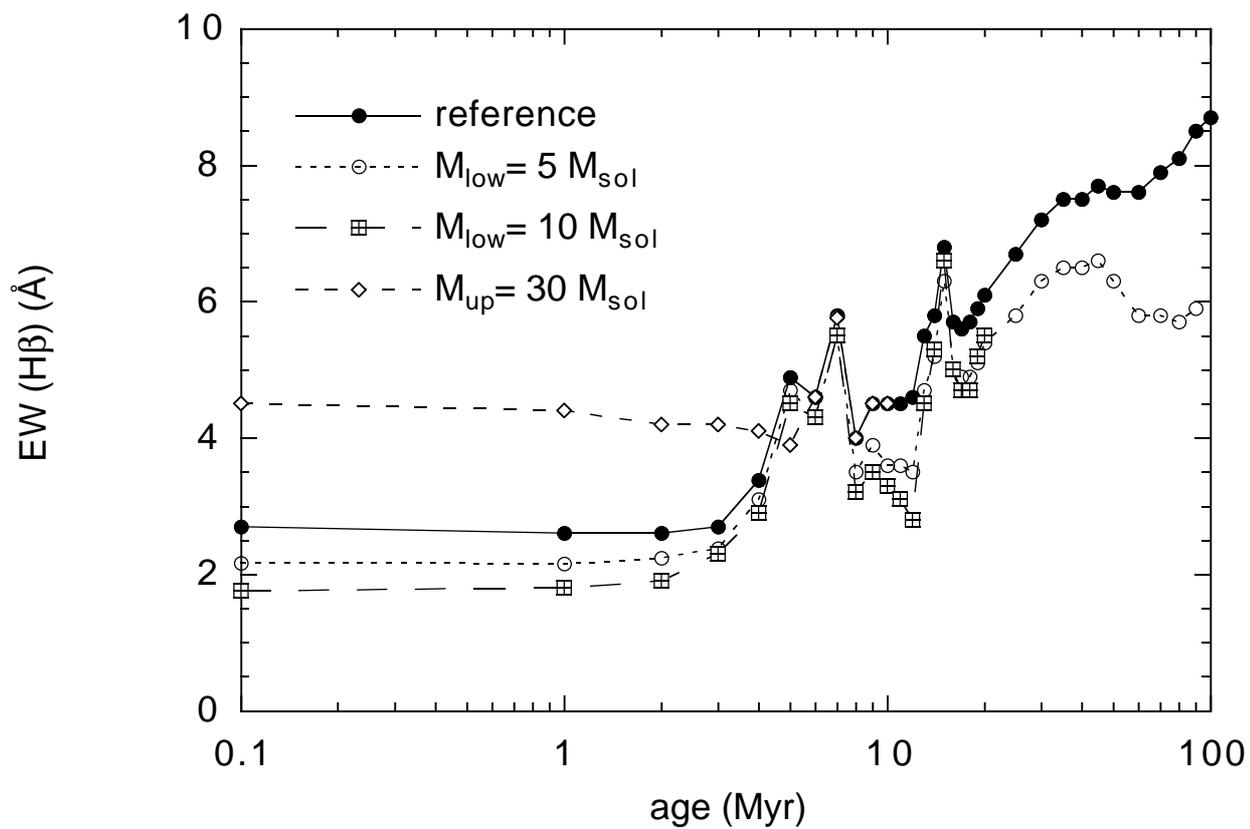
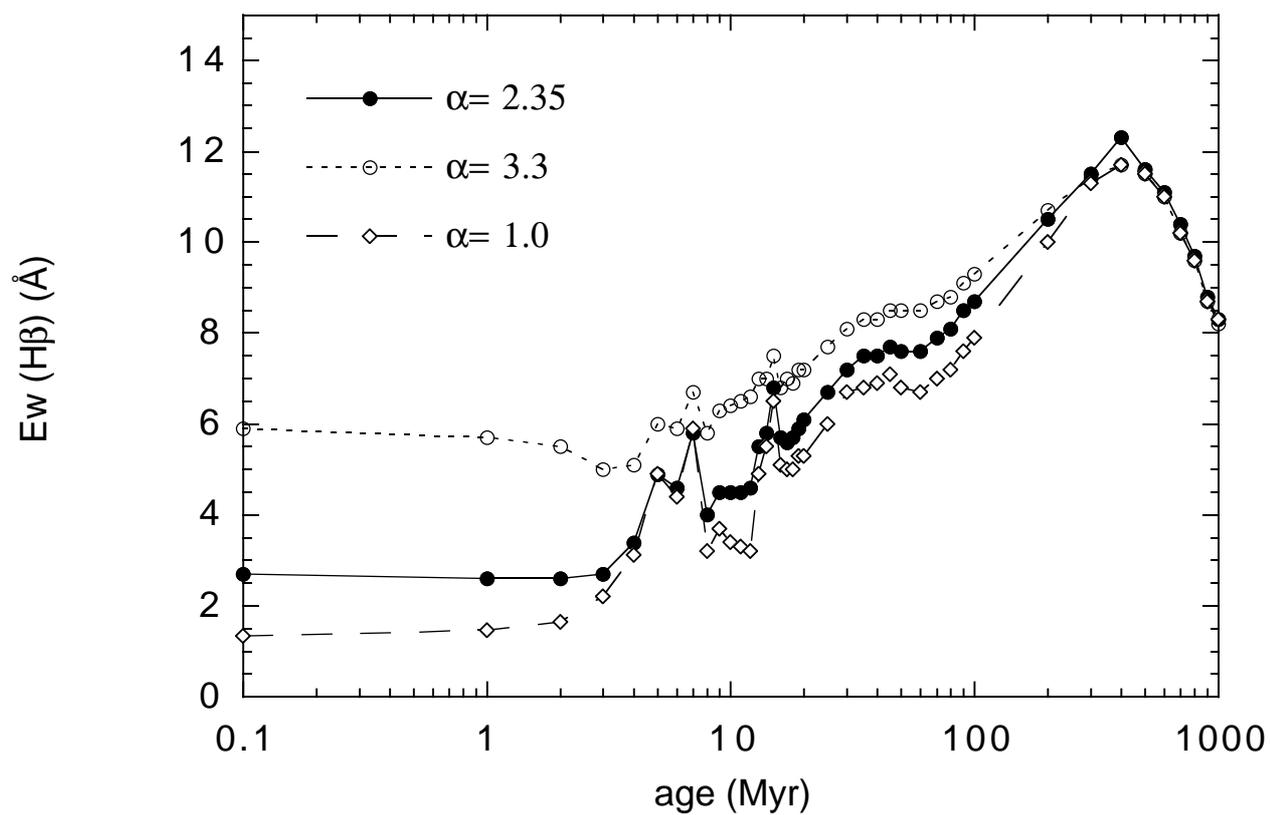
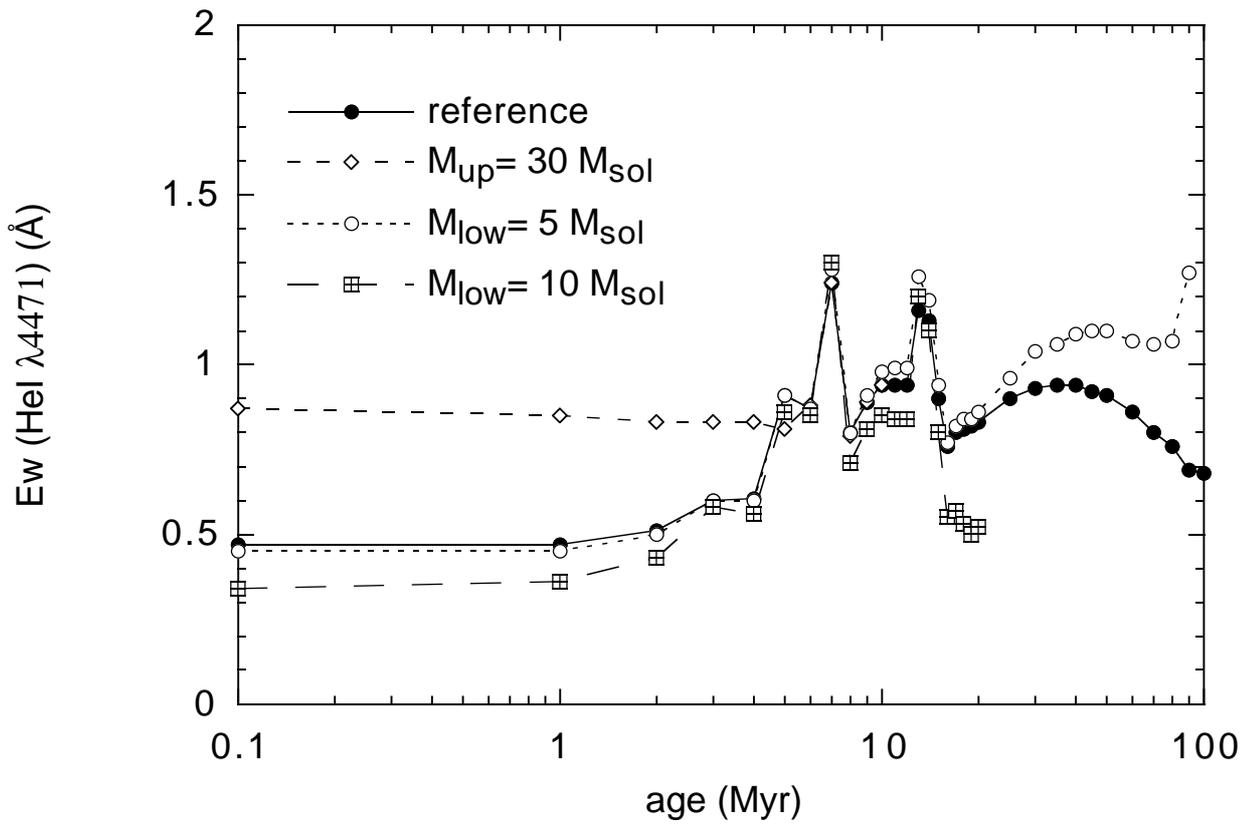
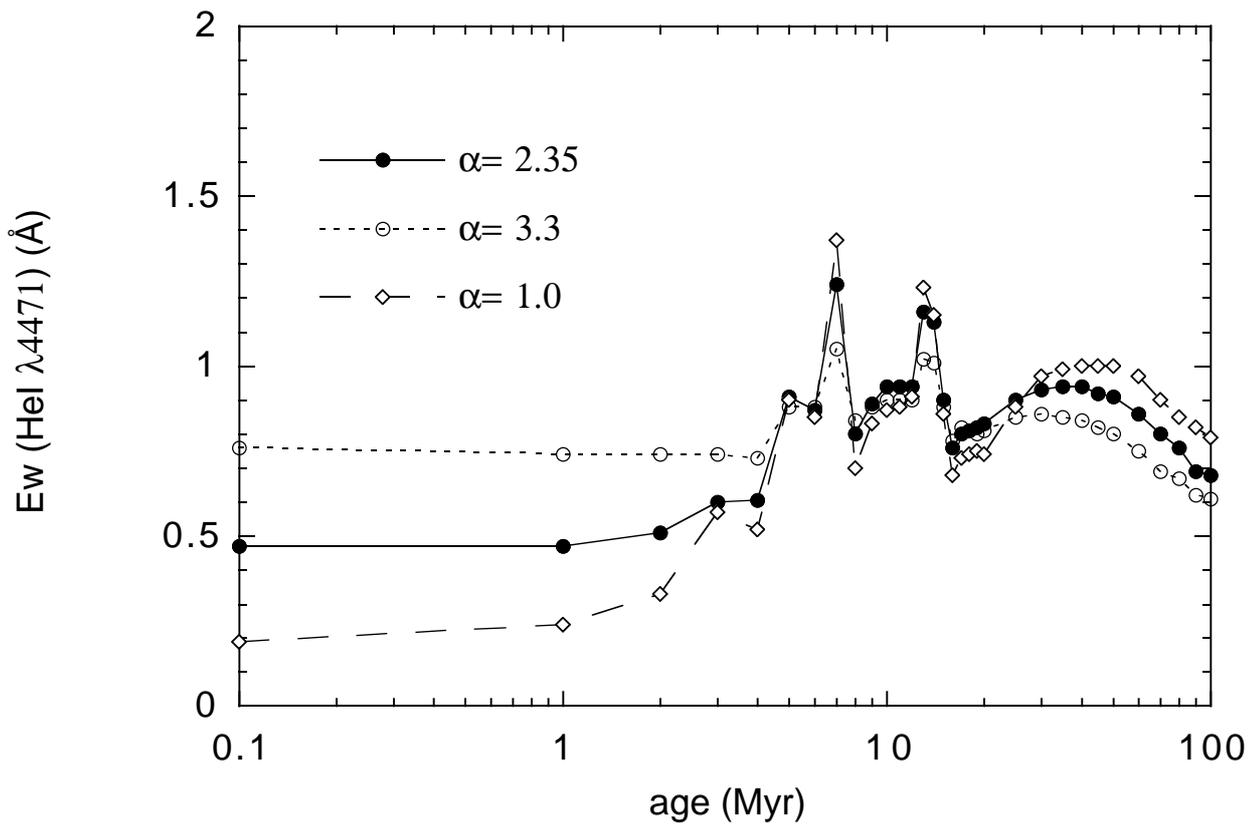
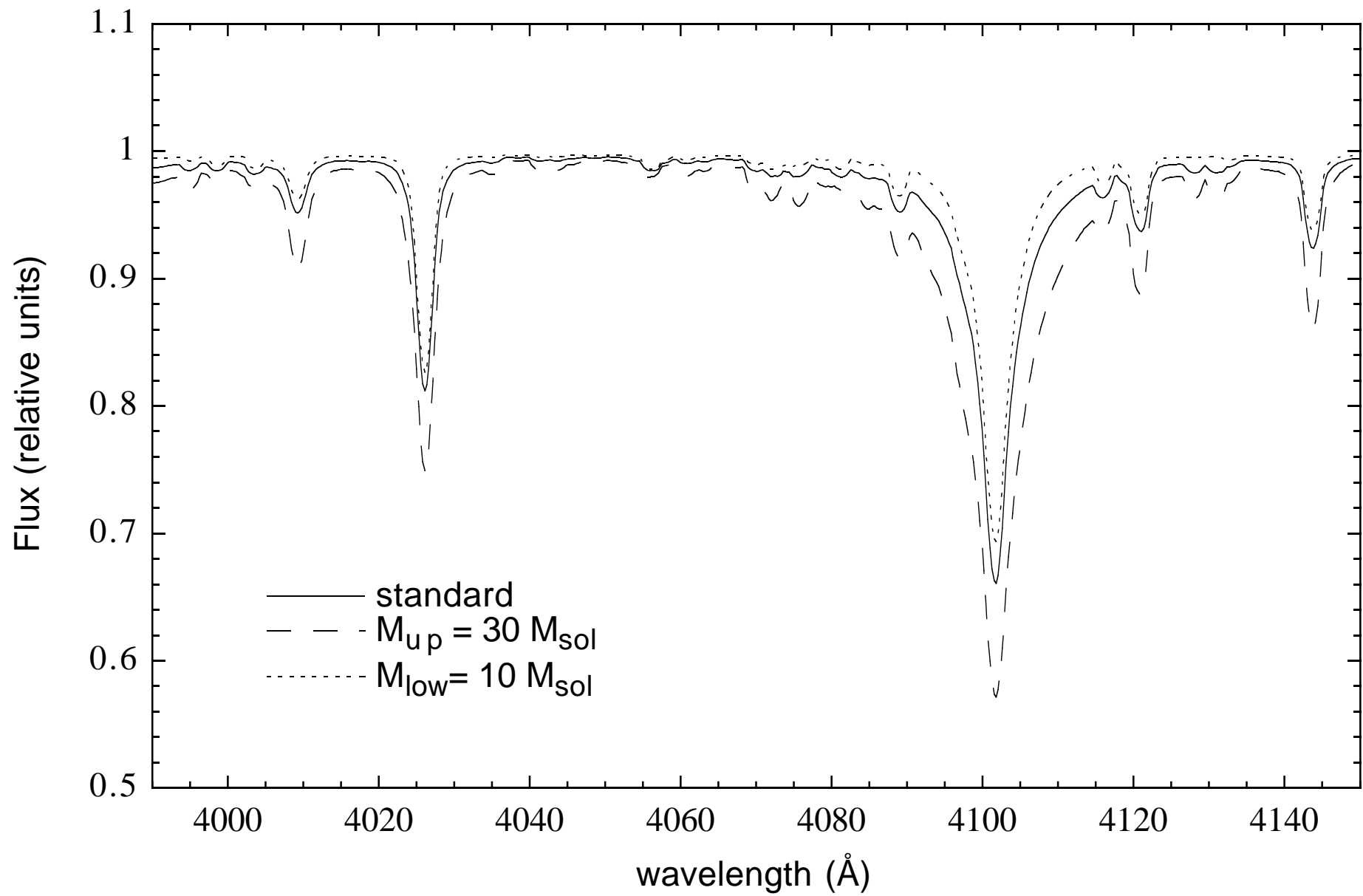
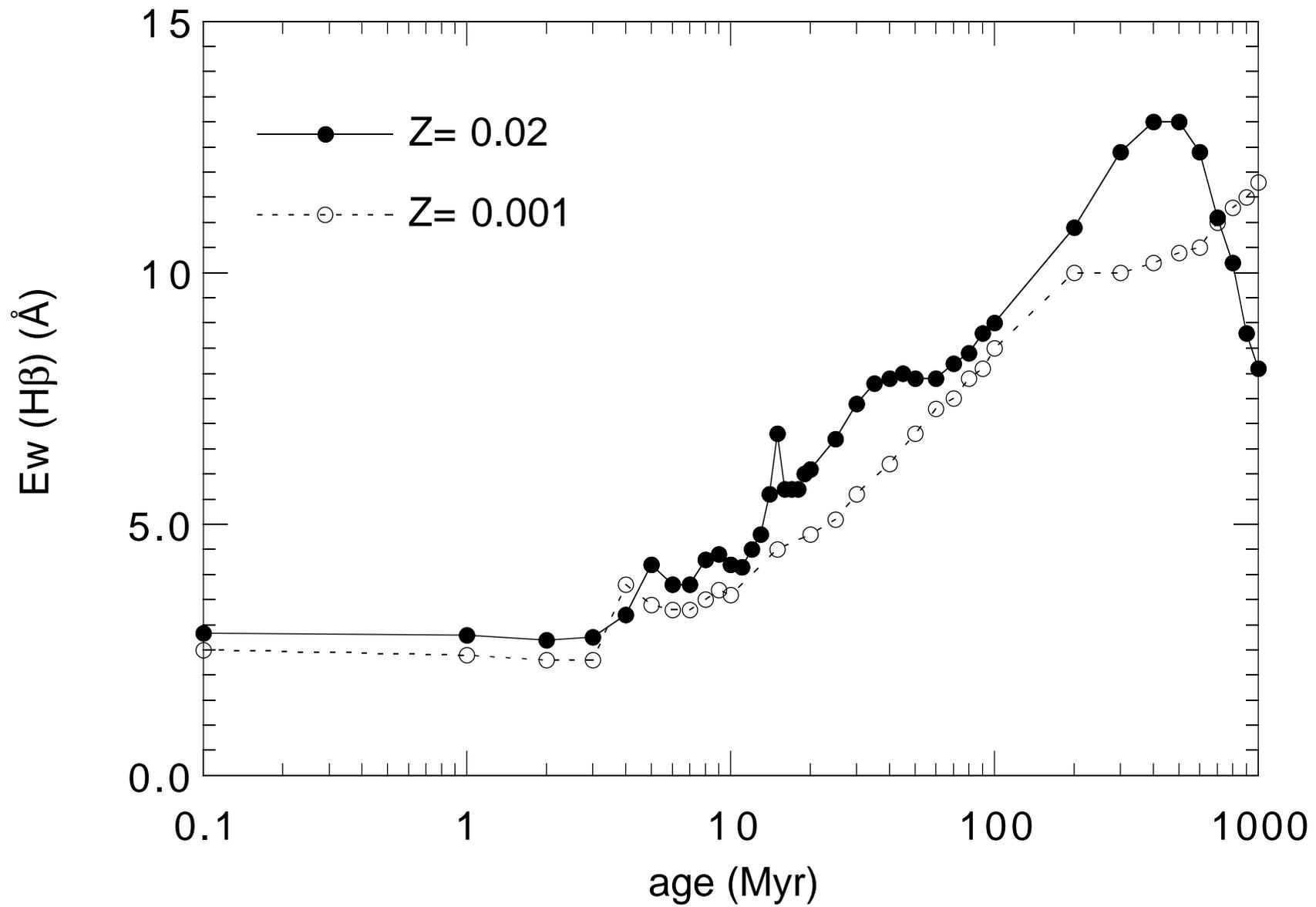
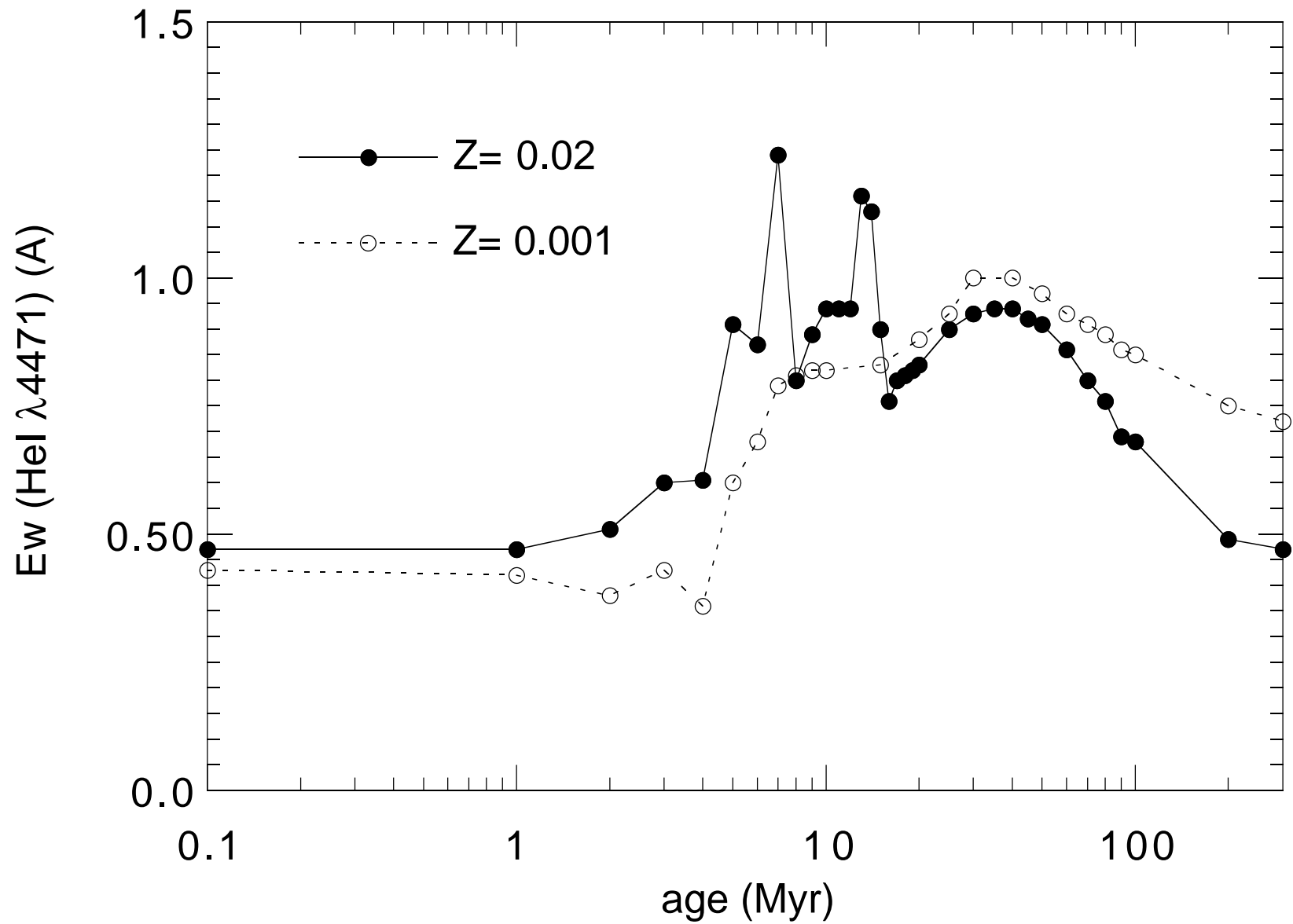
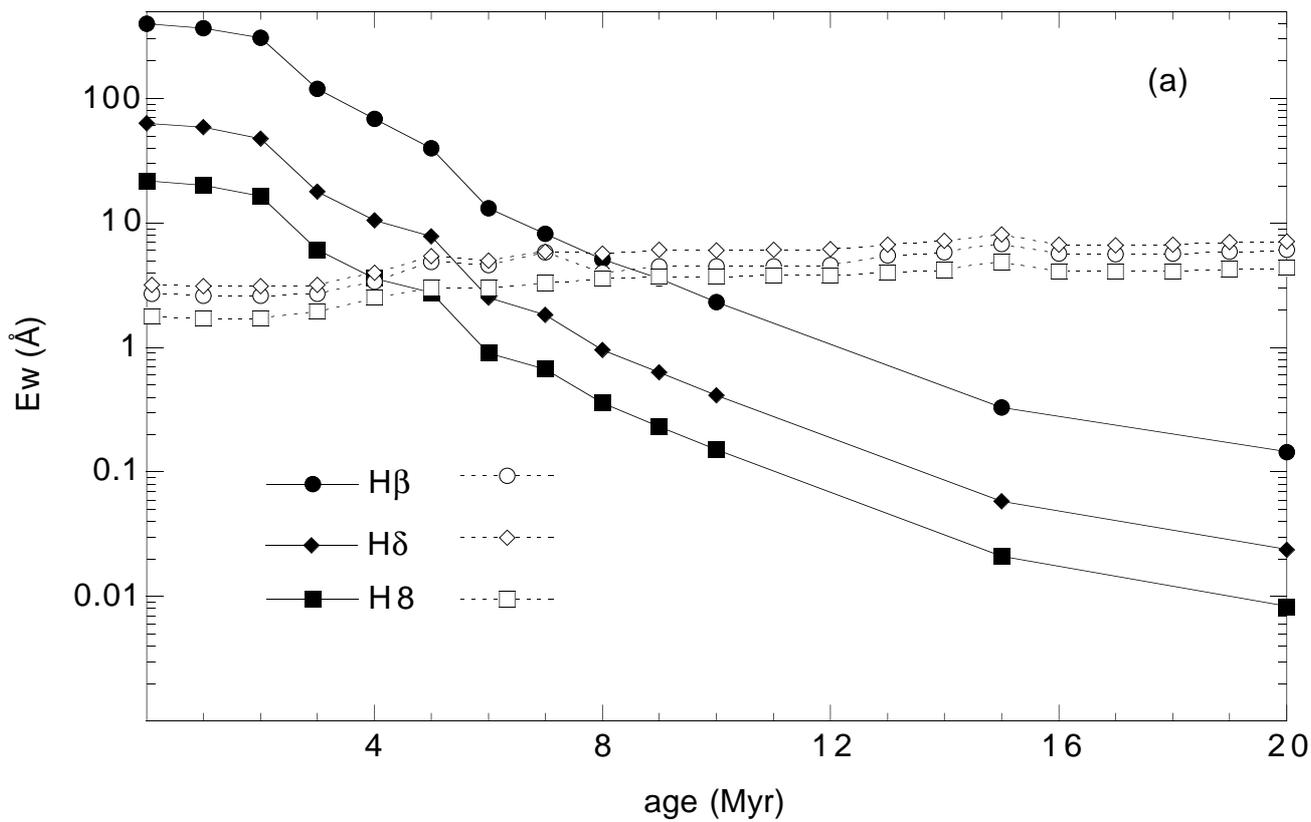
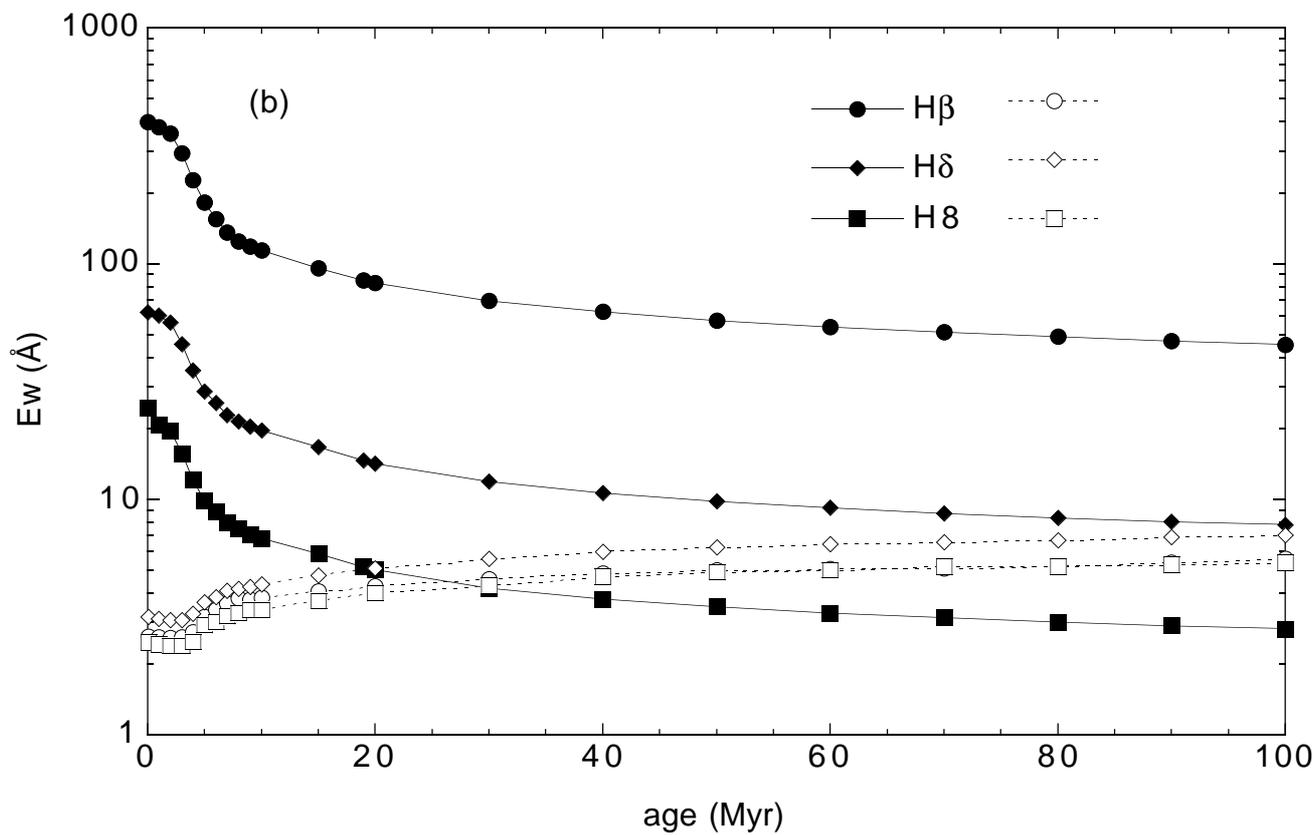
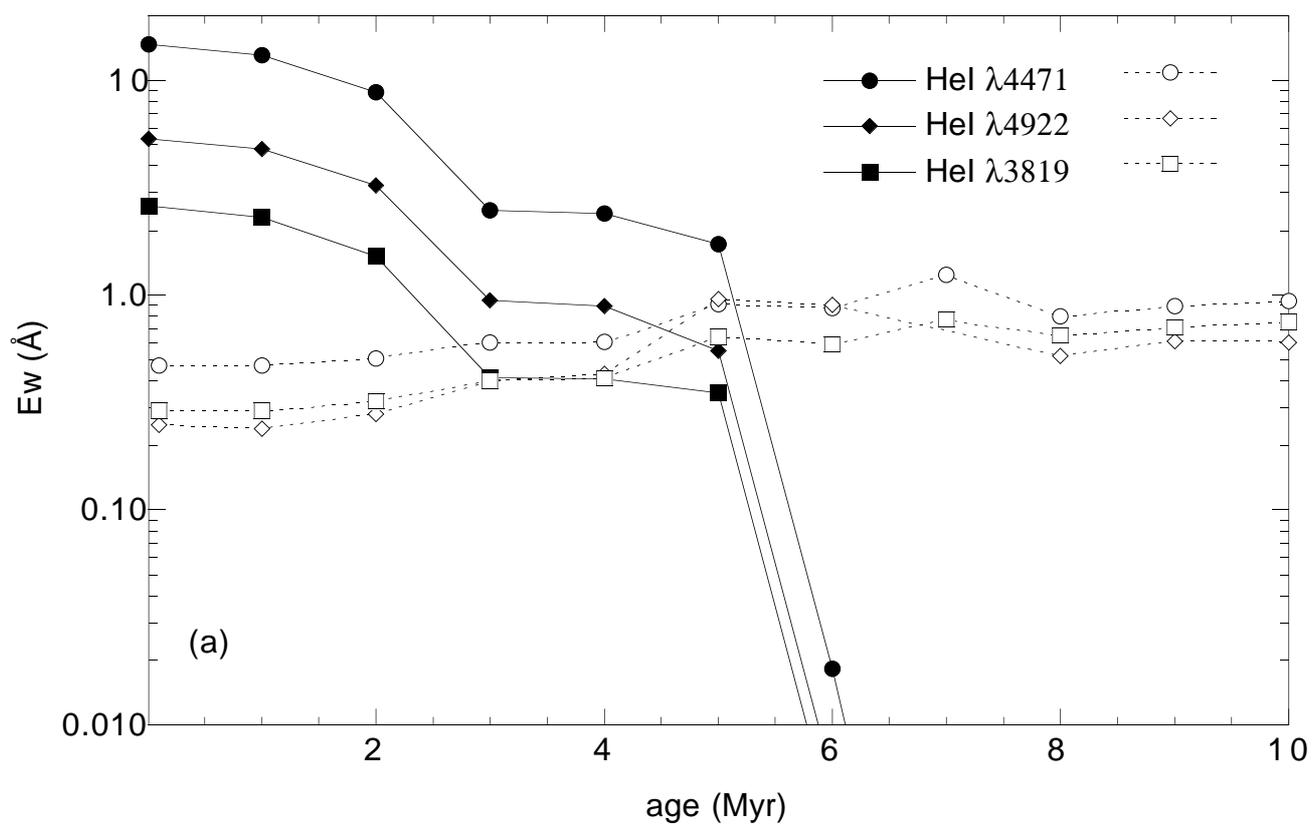
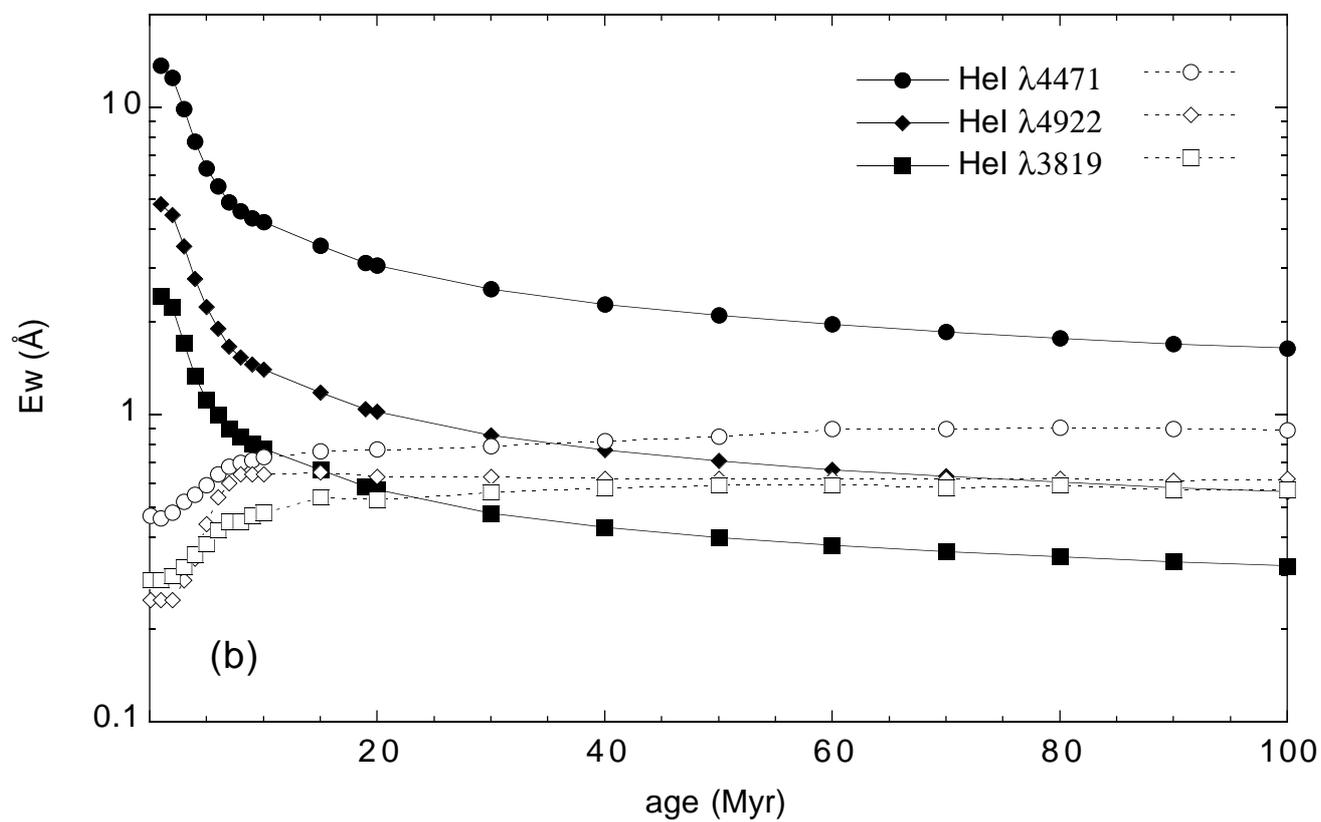
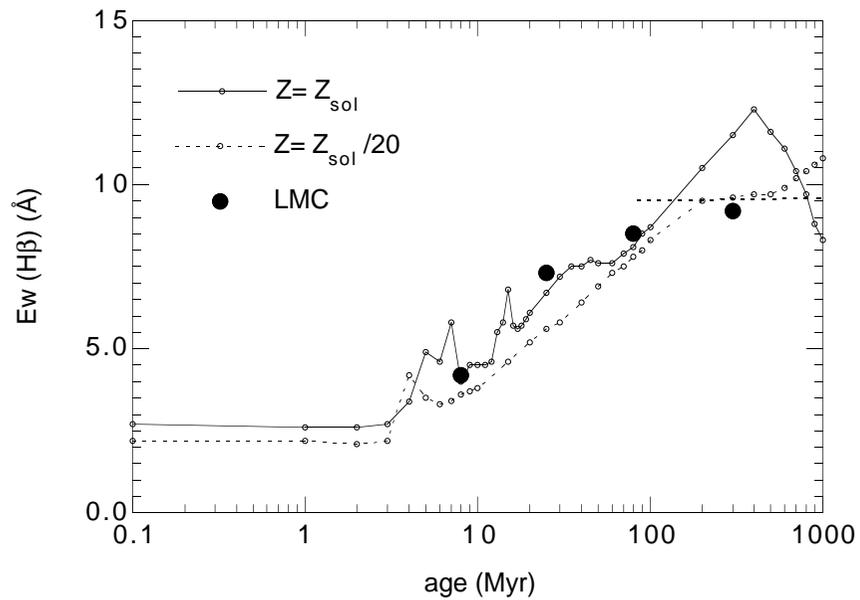
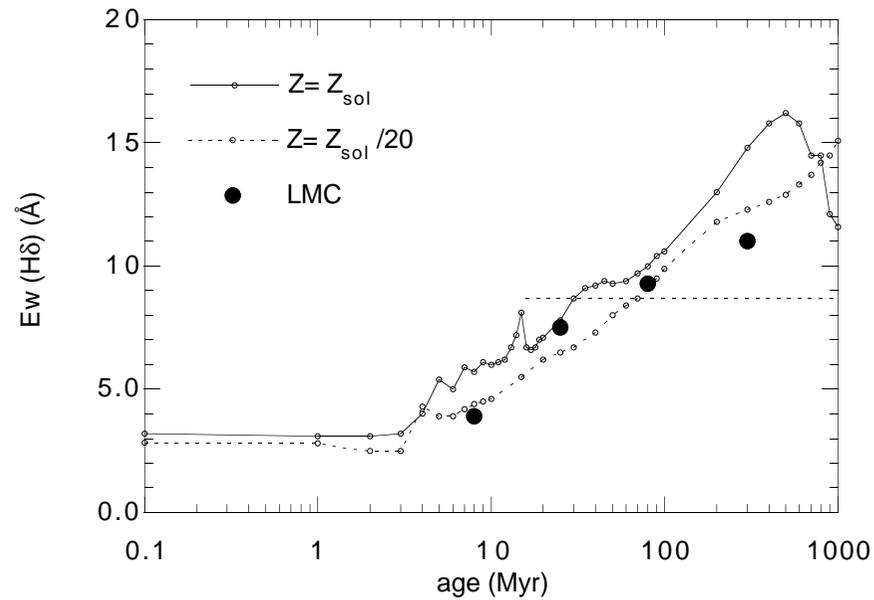
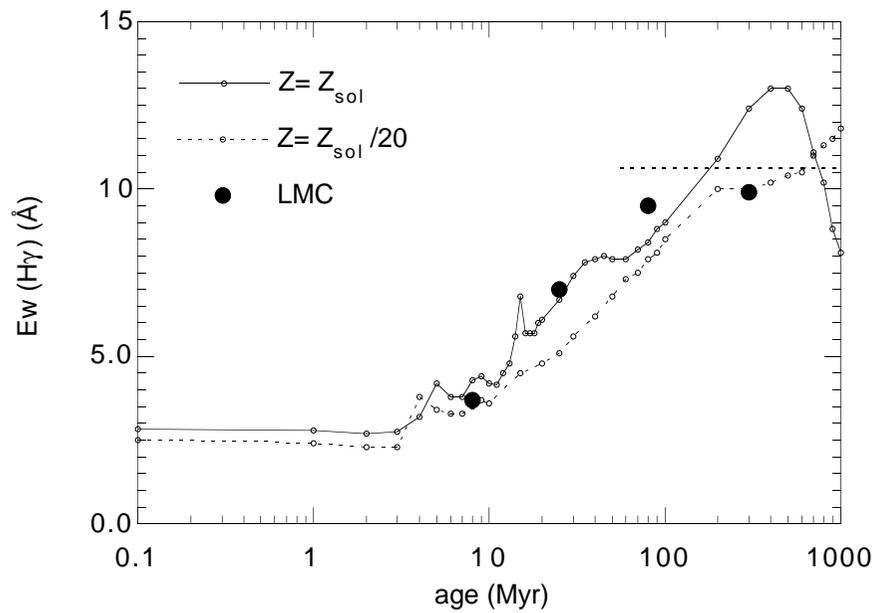
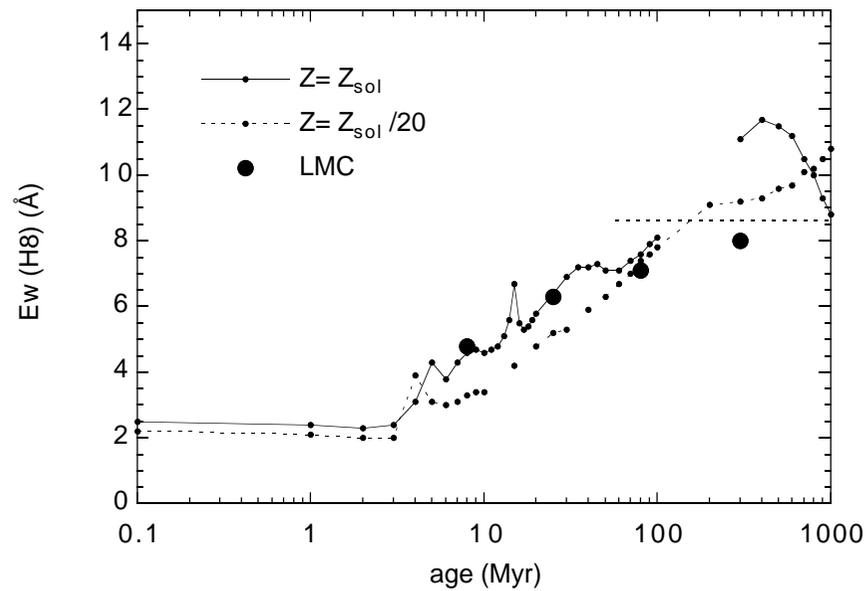
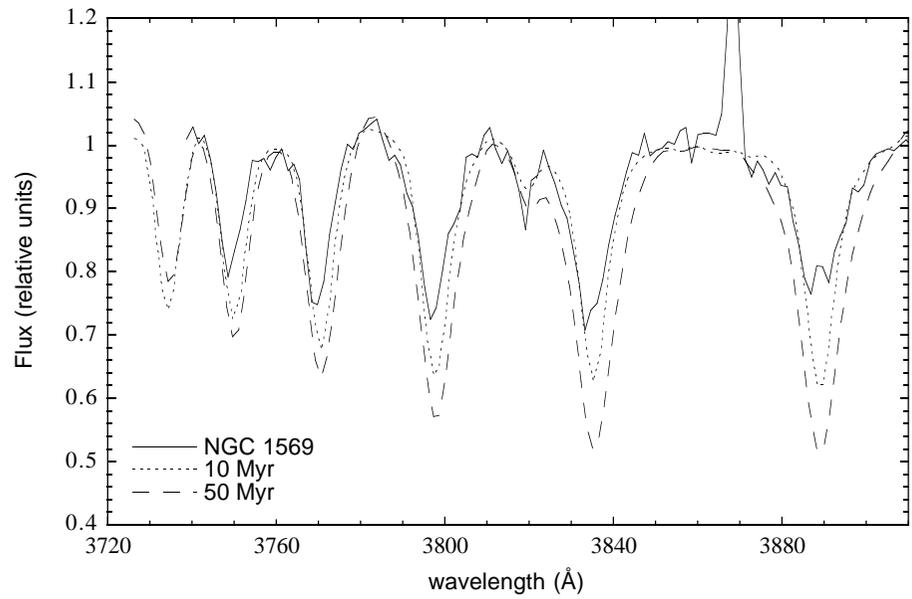
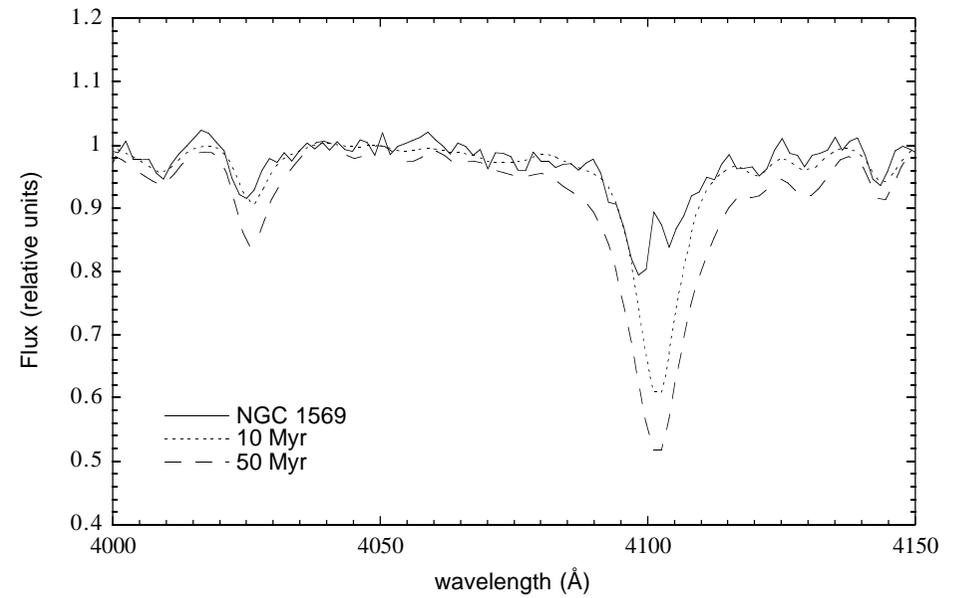
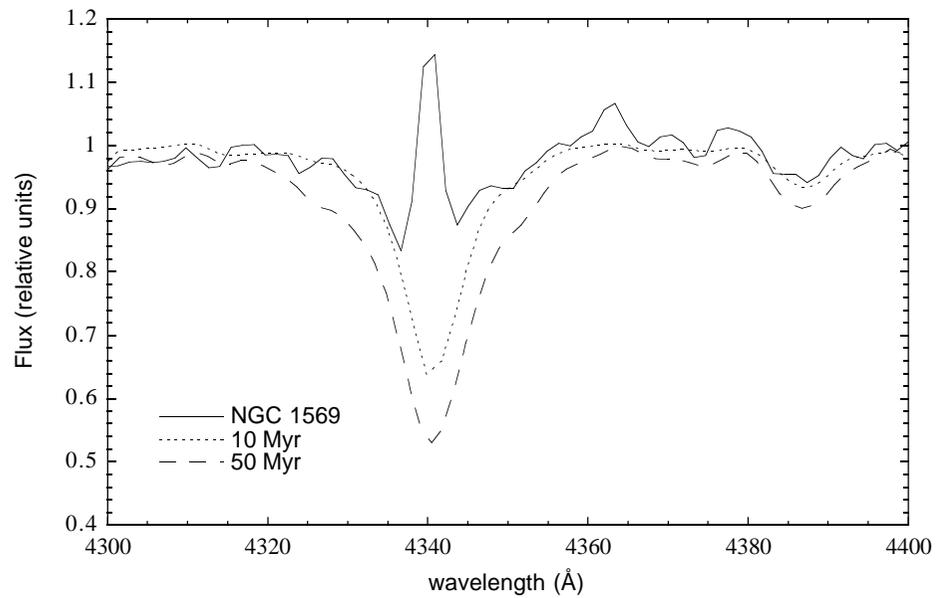
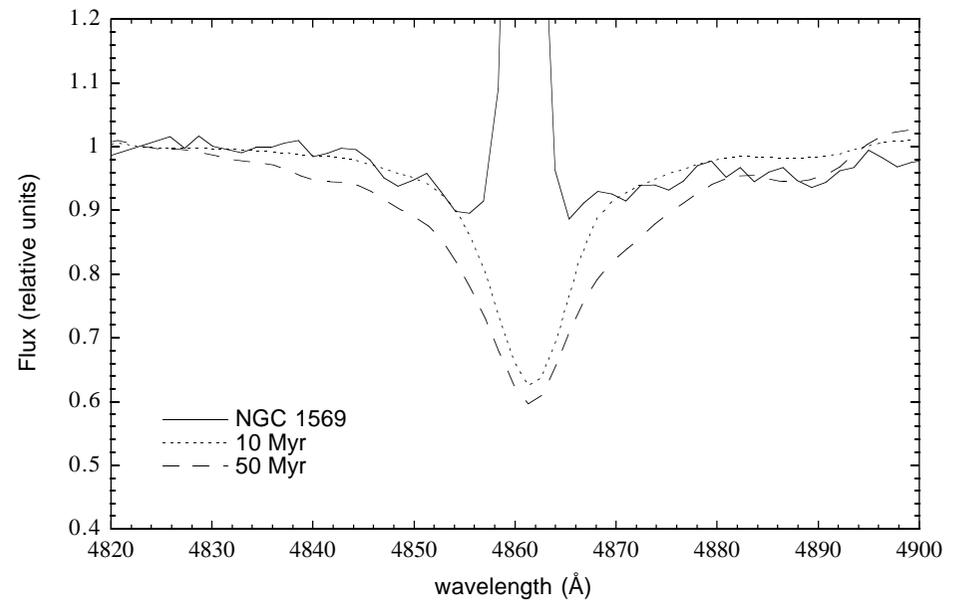
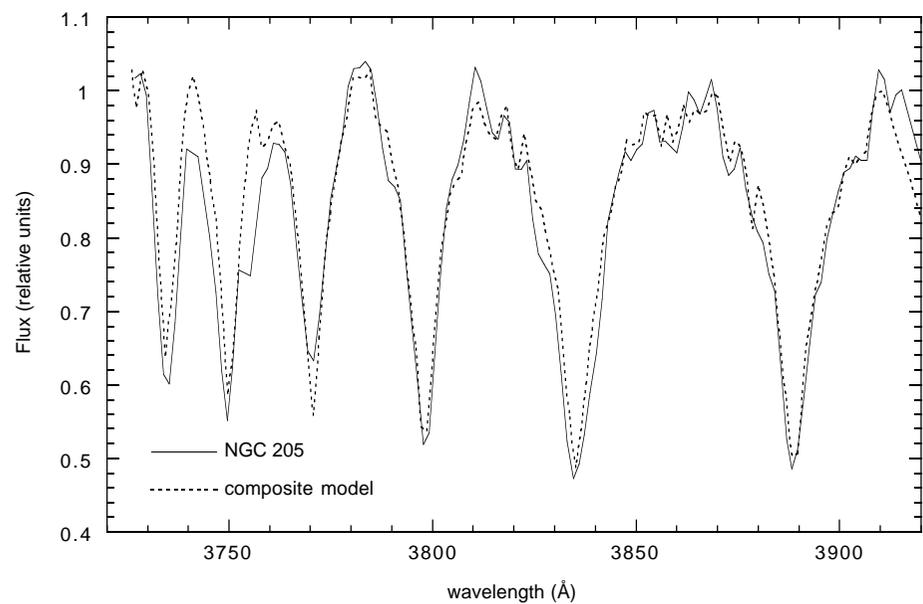
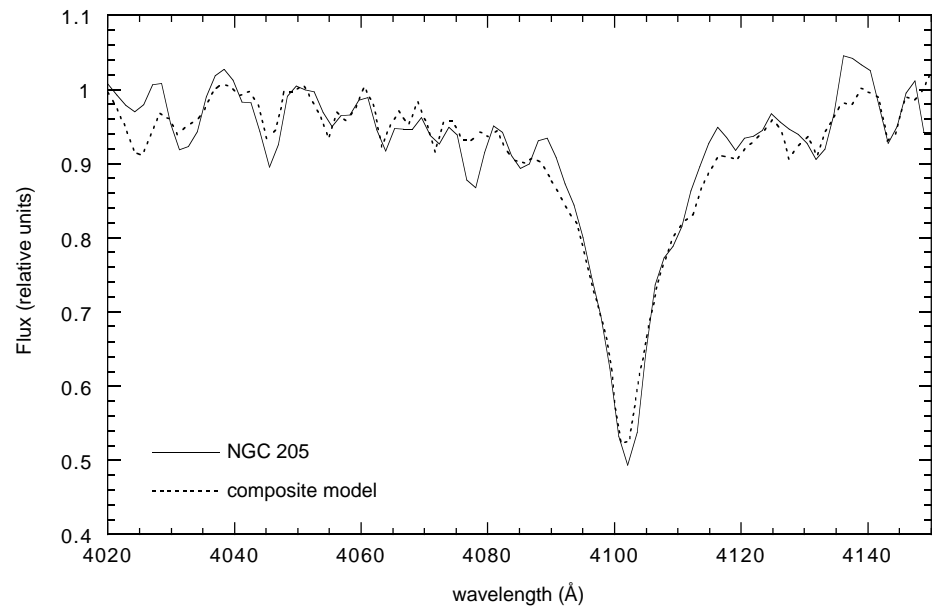
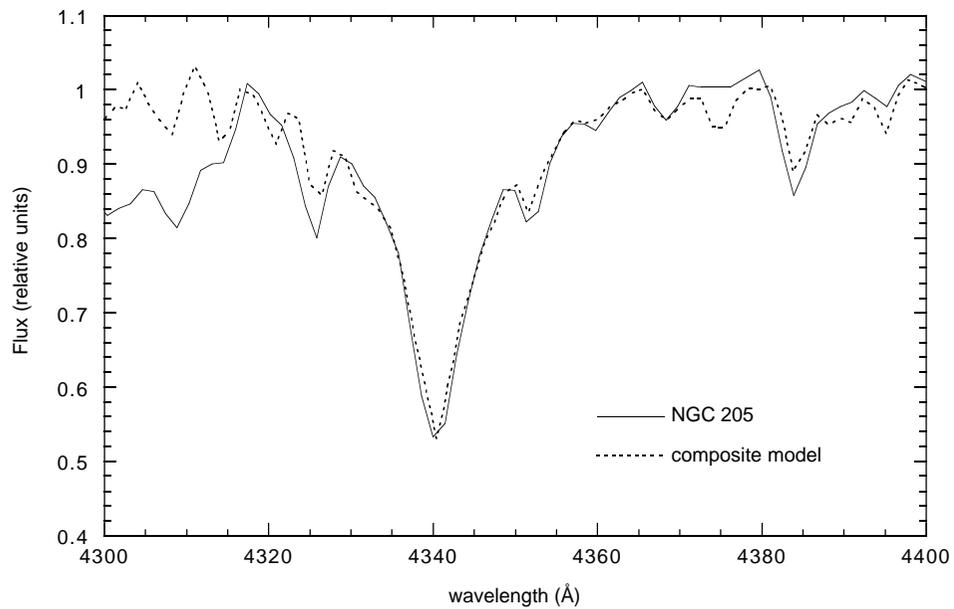
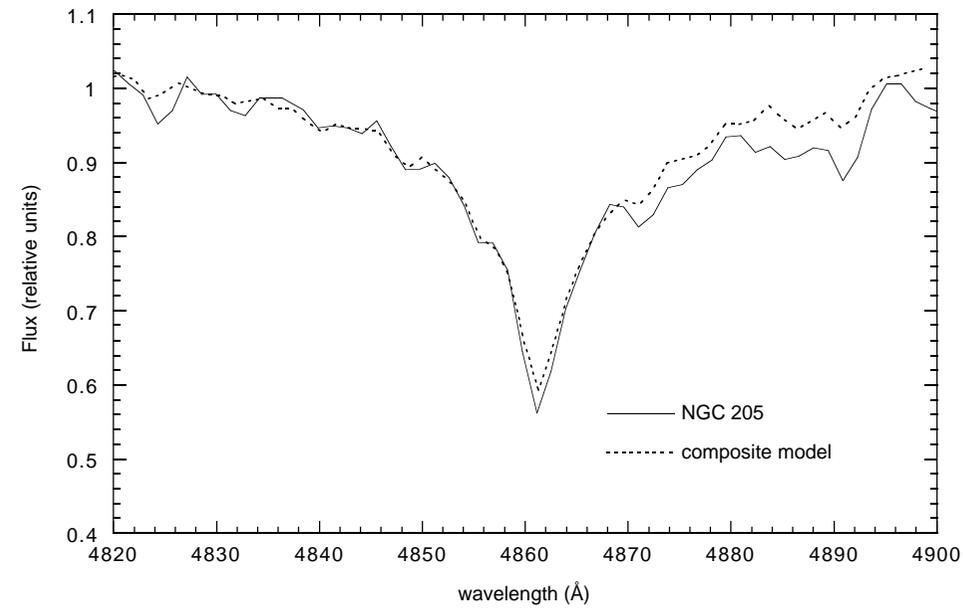
\end{document}